\documentclass[aps,twocolumn,superscriptaddress,prl,10pt,floatfix]{revtex4-2}

\usepackage[T1]{fontenc}
\usepackage{color}
\usepackage[english]{babel}
\usepackage{mathtools}
\usepackage{amsmath}
\usepackage{amssymb}
\usepackage{graphicx}
\usepackage[dvipsnames]{xcolor}
\usepackage{tikz-cd} 
\usepackage[justification=RaggedRight, singlelinecheck=false,figurename=FIG.]{caption} 
\usepackage{subcaption}
\usepackage{chngcntr}
\usepackage{svg}
\usepackage{esint}
\usepackage{tabularx,array, multirow}
\usepackage{comment}
\usepackage[font=small]{caption} 
\usepackage[unicode=true,pdfusetitle,bookmarks=true,bookmarksnumbered=false,bookmarksopen=false,
 breaklinks=false,pdfborder={0 0 1},backref=false,
 colorlinks = true, 
 citecolor = magenta,
 urlcolor = blue,
 linkcolor = red]
 {hyperref}
\usepackage{cleveref}
\crefname{figure}{Fig.}{Figs.}

\crefrangelabelformat{figure}{#3#1#4--#5\crefstripprefix{#1}{#2}#6}

\makeatletter

\newcolumntype{Y}{>{\centering\arraybackslash}X}
\newcommand{\phantomlabel}[2]{
    \protected@write\@auxout{}{
        \string\newlabel{#2}{
            {\@currentlabel#1}{\thepage}
            {\@currentlabel#1}{#2}{}
        }
    }
    \hypertarget{#2}{}
}

\DeclareFontFamily{OMX}{MnSymbolE}{}
\DeclareSymbolFont{MnLargeSymbols}{OMX}{MnSymbolE}{m}{n}
\SetSymbolFont{MnLargeSymbols}{bold}{OMX}{MnSymbolE}{b}{n}
\DeclareFontShape{OMX}{MnSymbolE}{m}{n}{
    <-6>  MnSymbolE5
   <6-7>  MnSymbolE6
   <7-8>  MnSymbolE7
   <8-9>  MnSymbolE8
   <9-10> MnSymbolE9
  <10-12> MnSymbolE10
  <12->   MnSymbolE12
}{}
\DeclareFontShape{OMX}{MnSymbolE}{b}{n}{
    <-6>  MnSymbolE-Bold5
   <6-7>  MnSymbolE-Bold6
   <7-8>  MnSymbolE-Bold7
   <8-9>  MnSymbolE-Bold8
   <9-10> MnSymbolE-Bold9
  <10-12> MnSymbolE-Bold10
  <12->   MnSymbolE-Bold12
}{}

\let\llangle\@undefined
\let\rrangle\@undefined
\DeclareMathDelimiter{\llangle}{\mathopen}%
                     {MnLargeSymbols}{'164}{MnLargeSymbols}{'164}
\DeclareMathDelimiter{\rrangle}{\mathclose}%
                     {MnLargeSymbols}{'171}{MnLargeSymbols}{'171}

\@ifundefined{showcaptionsetup}{}{%
 \PassOptionsToPackage{caption=false}{subfig}}

\renewcommand{\Im}{\text{Im}}

\newcommand{\roundbracket}[1]{\left(#1\right)}
\newcommand{\squarebracket}[1]{\left[#1\right]}

\begin{document}

\title{Light-Matter-Coupling formalism for magnons: probing quantum geometry with light}

\author{Ying Shing Liu}
\affiliation{Institute for Theoretical Solid State Physics, RWTH Aachen University, 52074 Aachen, Germany}

\author{Emil {Vi\~{n}as Bostr\"{o}m}}
\affiliation{Max Planck Institute for the Structure and Dynamics of Matter, Center for Free Electron Laser Science, 22761 Hamburg, Germany.}
\author{Michael A. Sentef}
\affiliation{Institute for Theoretical Physics and Bremen Center for Computational Materials Science, University of Bremen, 28359 Bremen, Germany}
\affiliation{Max Planck Institute for the Structure and Dynamics of Matter, Center for Free Electron Laser Science, 22761 Hamburg, Germany.}
\author{Silvia {Viola Kusminskiy}}
\email{kusminskiy@physik.rwth-aachen.de}
\affiliation{Institute for Theoretical Solid State Physics, RWTH Aachen University, 52074 Aachen, Germany}
\affiliation{Max Planck Institute for the Science of Light, Staudtstraße 2, 91058 Erlangen, Germany}

\date{\today}

%-----------------------------------------------------------------
% Abstract 
%-----------------------------------------------------------------

\begin{abstract} 
Nontrivial quantum geometry is a key feature of the wavefunctions of collective magnetic excitations in topological systems, but accessing it experimentally remains an open challenge. While Raman circular dichroism (RCD) has emerged as a promising probe, the fundamental link between the RCD and magnon quantum geometry has remained unsettled, and complicated by the fact that magnons are charge neutral. Here, we identify when and why this link exists. We show that, under broad conditions, the Fleury-Loudon Raman vertex can be obtained directly from a light-matter coupling expansion of the effective magnon Hamiltonian, bypassing the conventional microscopic derivation based on virtual electronic processes. This yields an analytical connection between the RCD and the Berry curvature of magnon bands. Applied to monolayer CrI\textsubscript{3}, our theory predicts finite temperature signatures of topological magnons in the RCD. These results establish a general route to quantum-geometry sensitive optical probes in magnonic systems.
\end{abstract}

\maketitle

%-----------------------------------------------------------------
% Introduction
%-----------------------------------------------------------------

\textit{Introduction}. Raman scattering is a versatile optical probe that non-invasively reveals the vibrational, structural, and magnetic properties of materials. This makes it an indispensable tool in condensed matter research, in particular for elucidating phonon and magnon dynamics, and their interplay~\cite{Loudon1964, Devereaux2007, Li2012, weber2013raman, Zhang2015, smith2019modern, Cui2023, Kumawat_2024}. In two-magnon Raman scattering, an optical photon is inelastically scattered by a magnetic material through the creation or annihilation of a pair of magnons, leading to the appearance of characteristic Stokes and anti-Stokes peaks at twice the single magnon frequency. Two-magnon Raman scattering can be used to probe e.g. antiferromagnets (AFM)~\cite{Elliott_1969, Cottam_1972, Lockwood-Cottam_1973, Lockwood1992, Lockwood-Cottam_2012} and spin liquids \cite{Knolle2014,yamamoto2020,Choi2021-oq}, giving insight into physical processes which cannot be probed via single-magnon scattering~\cite{Devereaux2007, Sahasrabudhe2020, Trebst2022, Calderon2015, Rigitano2020, ghosh_raman_2025}. For magnetic systems predicted to have topological magnon bands, where standard methods used in electronic systems to directly probe the quantum geometry~\cite{Klitzing1980, zhang_experimental_2005, Hasan_Kane_2010, Chang2013, Qi_Zhang_2011, ma_topology_2021, Torma_2023, kang_measurements_2025} are not available due to the lack of a Fermi energy, two-magnon Raman scattering in combination with the Raman circular dichroism (RCD) has emerged as a promising alternative~\cite{Bostroem2023}. Recently, this mechanism has also been proposed to probe the quantum geometric properties of unconventional magnets, such as chiral quantum spin liquids~\cite{koller2025ramancirculardichroismquantum} and altermagnets~\cite{yuan2025quantumgeometryaltermagneticmagnons}, as well as for probing spectral signatures of anyonic quasiparticles~\cite{Achim2025}. 

The Fleury-Loudon (FL) formalism is a well established framework to describe two-magnon Raman scattering~\cite{FleuryLoudon1968, ShastryShraiman1990, ShastryShraiman1991, Freitas2000}. The so-called FL Raman scattering vertex is governed by the exchange interaction, and can be derived microscopically from a Mott-Hubbard Hamiltonian coupled to light~\cite{Lee2010,Yang2021_nonFL}. The question arises whether it is possible to derive the same vertex directly from the effective magnon model, thereby bypassing complex light-matter interactions at the microscopic level, by using a method similar to the minimal coupling substitution for electronic systems: There, the coupling to light can be captured by the replacement ${\bf k}\rightarrow {\bf k}-e{\bf A}$ where ${\bf k}$ and $e$ are the electronic momentum and charge, and ${\bf A}$ is the vector potential~\cite{Michael2020_CD, Topp2021}. However, since magnons do not have an electric charge, it is not clear whether this is at all possible. Nonetheless, it was recently shown that such a substitution is possible for a particular system, namely a two-dimensional (2D) canted antiferromagnet on the honeycomb lattice~\cite{Bostroem2023, our_long_paper}. Further, the effective model has been applied successfully to other spin systems \cite{koller2025ramancirculardichroismquantum}. 

In this work, we show that the FL Raman vertex can be obtained directly from the magnon Hamiltonian, whenever this can be written in the presence of light as 
\begin{equation}
    H({\bf k},e{\bf A})= \frac{1}{2}\squarebracket{H({\bf k}-e{\bf A})+H({\bf k}+e{\bf A})}\, . \label{eq:even-combination-of-k-eA}
\end{equation}
We further provide the conditions under which Eq.~\eqref{eq:even-combination-of-k-eA} is valid, namely, for effective spin models arising from a microscopic Hubbard model where only direct hopping terms are considered, up to second order in a $t/U$ expansion. Under these conditions, optical observables associated with the FL Raman vertex can be immediately obtained in terms of light-matter couplings (LMCs)~\cite{Bostroem2023}, in direct analogy to the case of electronic systems~\cite{Passos_nlop_2018,Topp2021}. The LMCs are defined in terms of derivatives of the effective magnon Hamiltonian in the absence of light, reducing the complexity of the calculations enormously.

To illustrate the method, we apply our results to 2D honeycomb van der Waals ferromagnets (FM), describing a class of materials predicted to host topolological magnon bands. Examples include Cr$X$\textsubscript{3} ($X=$~I or Br)~\cite{Chen2018, Chen_2021_CrI3_not_K-Gamma, Cai_2021} and  Kitaev topological ferromagnets~Cr$X$Te\textsubscript{3} ($X =$~Si or Ge)~\cite{Zhu2021}. Using the LMC formalism to calculating the FL Raman vertex, we show that starting from a thermally populated state, the two-magnon RCD gives information on the underlying quantum geometry of the magnon bands. Specifically, we show that the RCD in momentum space is directly proportional to the Berry curvature, and that the contribution of the quantum metric is stored in a weighting function. The signal vanishes in the topologically trivial case.

%-----------------------------------------------------------------
% Fleury-Loudon vertex and the shortcut
%-----------------------------------------------------------------

\textit{Fleury-Loudon vertex and the shortcut}. The standard microscopic route to obtain the FL vertex starts (in the single-orbital case) from the Hubbard model
\begin{equation}
    H_{\text{micr}} = -\sum_{\langle ij\rangle \sigma} t_{ij}(c_{i\sigma}^\dagger c_{j\sigma}+\text{h.c.}) + U\sum_i n_{i\uparrow} n_{i\downarrow}. \label{eq:micro-Hubbard-Mott-model}
\end{equation}
The system is assumed to be in the Mott-insulator limit $t_{ij}\ll U$, with $t_{ij}$ the amplitude for an electron to hop between sites $i$ and $j$, and with $U$ the onsite Coulomb interaction. Here, $c_{i,\sigma}$ is the annihilation operator of an electrons with spin $\sigma$, and $n_{i,\sigma} =c_{i,\sigma}^\dagger c_{i,\sigma}$ is the number operator. Using second order perturbation theory in $t/U$, an effective spin Hamiltonian can be obtained, here generally denoted as $H^0$~\cite{MacDonald_tU_expansion_Hubbard_Mott, Yildirim1995, Takahashi_1977, Tasaki2020}. The light-matter interaction is introduced via Peierls substitution, amounting to electron hopping processes between different sites $c_{i\sigma}^\dagger c_{j\sigma}\rightarrow c_{i\sigma}^\dagger c_{j\sigma} e^{i\theta_{ij}(t)}$ acquiring a phase $\theta_{ij}(t)$~\cite{ShastryShraiman1990, ShastryShraiman1991, FleuryLoudon1968}, which is given by
\begin{equation}\label{eq:PeierlsP}
    \theta_{ij}(t)=-\frac{e}{\hbar}\int_j^i d\mathbf{r}\, {\bf A}({\bf r},t).
\end{equation}
By applying a Jordan map,
\begin{equation}
    c_{i\sigma}^\dagger c_{i\sigma'} = \frac{1}{2}n_i\delta_{\sigma'\sigma} +\roundbracket{{\bf S}_i \cdot \boldsymbol{\tau}}_{\sigma' \sigma },
\end{equation}
where $\boldsymbol{\tau}=(\tau_x,\tau_y,\tau_z)$ is the vector of Pauli matrices and $n_i=n_{i\uparrow}+n_{i\downarrow}$, the effective coupling Hamiltonian can be written in terms of spin operators.  A useful approach is to use a time-dependent Schrieffer-Wolff transformation~\cite{Eckstein2017}, which is a time-dependent perturbative approach to second order in $t/U$. 

For a Hamiltonian $H^0$ with only pairwise spin interactions, one obtains  
\begin{equation}
    H^{\text{c}}({\bf k},e{\bf A})=I_{ij}(t)H^0({\bf k})
    \label{Hceff}
\end{equation}
where we have written the result in Fourier ${\bf k}$-space and in the basis of the spin operators. Here
\begin{equation}
H^0 = \frac{1}{2}\sum_{\bf k}  H^0({\bf k}) {\bf S}_{\bf k}\cdot {\bf S}_{-\bf k}\,, 
\end{equation}
where ${\bf S}_{\bf k}$ is the Fourier transform of the local spin operators. The time-dependent coefficient reads
\begin{equation}
    I_{ij}(t) = \Im \int_{-\infty}^t dt' e^{i(U+i\eta)(t-t')}\cos\roundbracket{\theta_{ij}(t)-\theta_{ij}(t')} \label{eq:time-integral-peierls-phase},
\end{equation}
where $e^{-\eta(t-t')}$ with $\eta\rightarrow 0^+$ is a convergence factor. The FL vertex is obtained by second order perturbation theory in the vector potential~\cite{FleuryLoudon1968, ShastryShraiman1990, ShastryShraiman1991}. Using the dipole and rotating wave approximations, Eq.~\eqref{eq:time-integral-peierls-phase} is given to second order in ${\bf A}$ by
\begin{equation}
    I_{ij}(t)=\frac{e^2}{\hbar^2}\frac{1}{U-\omega_{\text{in}}}( {\bf A}_{\text{sc}}^*\cdot {\bf d}_{ij})( {\bf A}_{\text{in}}\cdot {\bf d}_{ij}) e^{i(\omega_{\text{in}}-\omega_{\text{sc}})t}. \label{eq:taylor-expanded-peierls-phase}
\end{equation}
Here ${\bf A}_{\text{in}}e^{i\omega_{\text{in}}t}$ (${\bf A}_{\text{sc}}e^{i\omega_{\text{sc}}t}$) is the vector potential field of the incident (scattered) photon with frequency $\omega_{\text{in}}$ ($\omega_{\text{sc}}$), and ${\bf d}_{ij}={\bf R}_i-{\bf R}_j$ with ${\bf R}_i$ lattice site $i$. Using $\omega_{\text{in}}-\omega_{\text{sc}}\ll\omega_{\text{in}},\,\omega_{\text{sc}}$ we can approximate $\omega_{\text{in}}\approx \omega_{\text{sc}}$ to obtain the standard FL Hamiltonian \cite{FleuryLoudon1968, ShastryShraiman1990, ShastryShraiman1991}
\begin{equation}
    H^{\text{FL}}({\bf k},e{\bf A}) = \frac{e^2}{\hbar^2}\frac{1}{U-\omega_{\text{in}}}({\bf A}_{\text{sc}}^*\cdot {\bf d}_{ij})({\bf A}_{\text{in}}\cdot{\bf d}_{ij}) H^0({\bf k})\,.
    \label{eq:standard-FL}
\end{equation}

As reported in Ref.~\cite{Yang2021_nonFL}, only direct electronic hopping processes lead to a vertex of the form of Eq.~\eqref{eq:standard-FL}. Other hopping processes in a multiorbital model, for example ligand-mediated hopping processes, generate additional terms. Those non-FL terms can be important for strong spin-orbit coupling materials, for instance, in Kitaev materials, where the Raman scattering intensity differs from the one obtained by the FL mechanism by at least two orders of magnitude~\cite{Yang2021_nonFL}. In this work we consider only the FL terms.

We note that if the spin Hamiltonian coupled to light (Eq.~\eqref{Hceff}) can be written as
\begin{equation}
    H^{\text{c}}({\bf k},e{\bf A})= \frac{1}{2}\squarebracket{H^0({\bf k}-e{\bf A})+H^0({\bf k}+e{\bf A})}\,, \label{eq:charge-neutral-k-A-dependence}
\end{equation}
Eq.~\eqref{eq:standard-FL} can be obtained by replacing ${\bf d}_{ij}$ by momentum derivatives,
\begin{equation}
 H^{\text{FL}}({\bf k},e{\bf A})=\frac{e^2}{\hbar^2}\frac{1}{U-\omega_{\text{in}}} ( {\bf A}_{\text{sc}}^*\cdot \nabla_{\bf k})( {\bf A}_{\text{in}}\cdot \nabla_{\bf k})H^0({\bf k})\,.
 \label{eq:FL_momentum_derivative}
\end{equation}
In the Supplementary Material (SM)~\cite{supplementary_material}, making use of the results from Refs.~\cite{Kitamura2017, Yildirim1995, Eckstein2017}, we show that as long as only direct hopping terms are considered, Eq.~\eqref{eq:charge-neutral-k-A-dependence} is valid to all orders in the vector potential for a second order expansion in $t/U$, consistent with the degree of approximation that we are using in this work. For higher orders in $t/U$, Eq.~\eqref{eq:charge-neutral-k-A-dependence} breaks down due to concatenated virtual hopping processes which do not fall on a straight line, in particular by the magnetic flux enclosed by closed-loop hopping processes (see SM~\cite{supplementary_material}). Our results are consistent with previous studies in the static field limit \cite{Diptiman_Chitra_spin_chirality_1995, Montrunich_spin_chirality_2006}.

Therefore, to second order in $t/U$, we can define light-matter couplings (LMCs) in a way analogous to that of electronic systems \cite{Topp2021}
\begin{equation}
H^{\text{c}}({\bf k},e{\bf A})= H^0({\bf k})+ \sum_{n=1} \sum_{\mu\cdots \nu} L_{\mu\cdots \nu}^{(n)}({\bf k}) (eA^\mu) \cdots (eA^\nu) .
 \end{equation}
Here $A^\mu$ is the $\mu^\mathrm{th}$ component of vector potential, and we have defined the $n^\mathrm{th}$ order LMC as
\begin{equation}
    L_{\mu\dots \nu}^{(n)}({\bf k}) = \frac{1}{n!}\underbrace{\partial_{\{eA_\mu\}} \dots \partial_{\{eA_\nu\}}}_{n} H^{\text{c}}({\bf k},e{\bf A})\vert_{{\bf A}=0}.
\end{equation}
For Hamiltonians of the form of Eq.~\eqref{eq:charge-neutral-k-A-dependence}, only terms of even order in ${\bf A}$ are present in the expansion, such that for even-order LMCs we can replace the vector potential derivatives with momentum derivatives,  
\begin{equation}
    L_{\mu\dots \nu}^{(2n)}({\bf k}) = \frac{1}{2n!}\underbrace{\partial_{\mu} \dots \partial_{\nu}}_{n} H^0({\bf k})\,.
    \label{eq:goodLMCs}
\end{equation}
Here we have introduced the notation $\partial_{k_\mu}\rightarrow\partial_\mu$. Relevant optical observables resulting from the scattering of an even number of photons can be obtained from these LMCs. In particular, the FL vertex in terms of the LMCs simply reads
\begin{equation}
   H^{\text{FL}}({\bf k}, e{\bf A})= \frac{2e^2}{\hbar^2(U-\omega_{\text{in}})} \sum_{\mu\nu} A_{\text{sc}}^{*\mu} A_{\text{in}}^\nu L_{\mu\nu}^{(2)}({\bf k}).
   \label{eq:FL_from_LMCs}
\end{equation} 
We therefore conclude that, to second order in $t/U$, the FL vertex can be obtained by means of a minimal-substitution like transformation, at the level of the effective spin Hamiltonian. 

%-----------------------------------------------------------------
% Case study: 2D honeycomb ferromagnets
%-----------------------------------------------------------------

\begin{figure}[!ht]
    \centering
    \includegraphics[width=\linewidth]{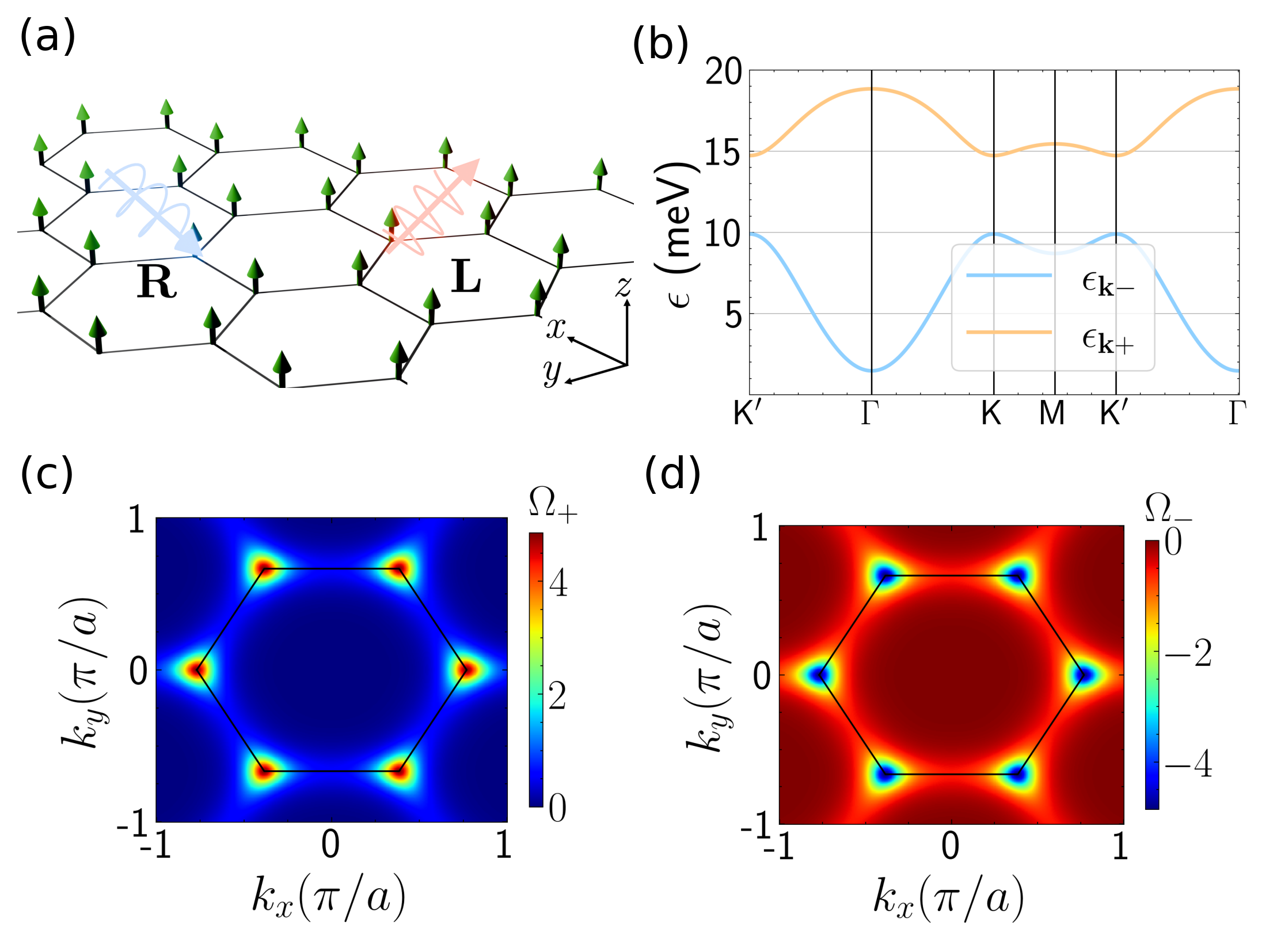}
    \begin{subfigure}{1\textwidth} % this 'subfigure' env. has no visible content
        \refstepcounter{subfigure}\label{subfig:2D_CrI3_set_up}
        \refstepcounter{subfigure}\label{subfig:CrI3_band_structure}
        \refstepcounter{subfigure}\label{subfig:CrI3_berry_curvature_upper}
        \refstepcounter{subfigure}\label{subfig:CrI3_berry_curvature_lower}
    \end{subfigure}%
    \vspace{-5mm}
    \caption{(a) Setup: incident circularly polarized light is scattered from a two-dimensional honeycomb ferromagnet. (b) Upper and lower magnon bands calculated from Eq.~\eqref{eq:free-magnon-Hamiltonian-second-quantized}. (c,d) Berry curvatures of the upper (c) magnon band and the lower (d) magnon band. Parameters are taken from Ref.~\cite{Chen2018}, and are $J_1 = 2.01$ meV, $J_2 = 0.16$ meV, $J_3 = -0.08$ meV, $A_z = 0.49$ meV, and $D = 0.31$ meV. The length of the spin vector is $S = 3/2$.}
    \label{fig:CrI3_set_up}
\end{figure}

\textit{Case study: 2D honeycomb ferromagnets}. We now apply the LMC formalism to compute the two-magnon RCD for a 2D honeycomb ferromagnet, see Fig.~\ref{subfig:2D_CrI3_set_up}. We show that a non-vanishing circular dichroism signal indicates a nontrivial topology of the magnon bands, similar to the result reported in Ref.~\cite{Bostroem2023} for a canted AFM on the honeycomb lattice. Differently from Ref.~\cite{Bostroem2023}, the relevant RCD processes for the FM are vertical interband transitions, which are activated at finite temperature. The temperature dependence of the two-magnon RCD signal serves to pinpoint the topological nature of the magnon bands.  

Using a Holstein-Primakoff (HP) transformation for the spin operators, the magnon Hamiltonian can be obtained from the spin Hamiltonian~\cite{HP1940, Chen2018}
\begin{align}\label{eq:H0_FM_SS}
    H^0 = &-\sum_{r=1}^3\sum_{\langle i,j\rangle_r} J_r {\bf S}_i\cdot {\bf S}_j - \sum_{\langle i,j\rangle_2} {\bf D}_{ij} \cdot ({\bf S}_i\times {\bf S}_j) \\ &-\sum_{i} A_z (S_i^z)^2. \nonumber
\end{align}
Here $\langle i,j\rangle_r$ denotes a sum over sites that are $r^\mathrm{th}$ nearest neighbors (n.n.), $J_r$ is the $r^\mathrm{th}$ n.n. exchange coefficient, ${\bf D}_{ij}$ is the DMI vector between sites $i$ and $j$, and $A_z$ is an easy-axis anisotropy.

%-----------------------------------------------------------------
% Figure: CrI3, its magnon bands and Berry curvature
%-----------------------------------------------------------------
 
One can show that both the exchange and DMI interactions satisfy Eq.~\eqref{eq:charge-neutral-k-A-dependence} to second order in $t/U$~\cite{supplementary_material}. The anisotropy term plays no role in the ${\bf k}$- or ${\bf A}$-dependence. 
Hence, we can apply the LMCs formalism to Eq.~\eqref{eq:H0_FM_SS}. In the spin-wave limit, Eq.~\eqref{eq:H0_FM_SS} reduces to a Hamiltonian of the form~\cite{Owerre2016, Chen2018, supplementary_material}
\begin{align}
     H^0 &\approx \frac{1}{2}\sum_{\bf k} \psi_{\bf k}^\dagger H^0_\text{M}({\bf k})\psi_{\bf k}\,. \label{eq:free-magnon-Hamiltonian-second-quantized} 
\end{align}
The basis of bosonic operators is given by $\psi_{\bf k}^\dagger = (a_{\bf k}^\dagger ,b_{\bf k}^\dagger)$ where $a_{\bf k}^\dagger\approx S_{A,-\bf k}^-/\sqrt{2S}$ and $a_{\bf k}^\dagger a_{\bf k} \approx S- S_{A,\bf k}^z $ for operators on sub-lattice $A$, and similarly for the operators on sub-lattice $B$. Here $S$ is the spin quantum number of the magnetic ion. The two-magnon FL Raman scattering vertex can be now obtained using Eqs.~\eqref{eq:goodLMCs} and~\eqref{eq:FL_from_LMCs}. The setup, calculated magnon bands and associated Berry curvature are all illustrated in Fig.~\ref{fig:CrI3_set_up}.

Let us now consider for definiteness the scattering of light at normal incidence, for a 2D sample lying in the $xy$-plane. We define the basis of circular polarization as ${\bf e}_{s} = \frac{1}{\sqrt{2}} (e_x+ise_y)$, with $s=\pm 1$, and $e_z$ is the propagation direction. 
Using Eq.~\eqref{eq:FL_from_LMCs} we obtain the two-magnon Raman scattering Hamiltonian
\begin{equation}
 H_R^{ss'} = \frac{1}{2}\sum_{\bf k} \psi_{\bf k}^\dagger H_{R,\bf k}^{ss'} \psi_{\bf k},
\end{equation}
where $s$ ($s'$) labels the incident (scattered) polarization of light, and
\begin{equation}
    H_{R,\bf k}^{ss'} = \frac{K_0}{2}\squarebracket{L_{xx}^{(2)}({\bf k})+ss'L_{yy}^{(2)}({\bf k}) +i(s-s')L_{xy}^{(2)}({\bf k})}. \label{eq:FL-vertex-in-LMCs}
\end{equation}
Here $K_0 = 2e^2 A_{\text{sc}} A_{\text{in}}/[\hbar^2(U-\omega_{\text{in}})]$, and the explicit form of $L_{\mu\nu}^{(2)}$ is obtained by applying Eq.~\eqref{eq:goodLMCs} to $H^0_{\text{M}}({\bf k})$. These expressions are given in the SM~\cite{supplementary_material}.

As we see in Fig.~\ref{fig:CrI3_set_up}, the Berry curvature is maximal at the edges of the Brillouin zone (BZ). Hence, to probe it, we need Raman processes covering the full BZ. Due to conservation of linear momentum, only vertical transitions are allowed. At finite temperature, the thermal population activates processes at finite ${\bf k}$. At a given temperature $T$, the Raman scattering cross-section is \cite{Bacci_Gagliano_1991}
\begin{equation}
 \mathcal{P}^{ss'}(\omega,T) = \sum_{mn} \frac{e^{-\beta E_m}}{Z} |\langle \Psi_{n} | H_{\rm R}^{ss'}|\Psi_{m}\rangle |^2 \delta(\hbar\omega - E_{nm}), 
\end{equation}
where $\beta = (k_B T)^{-1}$ is the inverse of the temperature $T$, $\omega=\omega_{\text{sc}}-\omega_{\text{in}}$ is the frequency difference between scattered and incident light, and $E_{nm} = E_n - E_m$ is the transition energy from the initial state to the final state. The initial states $|\Psi_{m}\rangle$ are weighted by the Boltzmann factor $e^{-\beta E_m}$, and the partition function is $Z = \sum_m e^{-\beta E_m}$. Since the FL vertex is diagonal in ${\bf k}$, we can label the eigenstates by $|\Psi_{m}\rangle = |m_{\bf k\alpha}\rangle$, where $|m_{\bf k\alpha}\rangle$ contains $m$ magnons with momentum ${\bf k}$ in band $\alpha$. In the following discussion, we restrict our attention to processes in which a magnon undergoes an interband transition from the lower, thermally occupied band, to the upper magnon band. We assume that the upper magnon band is initially unoccupied, which is usually valid for temperatures below the Curie temperature~\cite{cenker_direct_2021, huang_layer-dependent_2017}.
The Raman cross-section can then be written as
\begin{align}
 \mathcal{P}^{ss'}(\omega,T) &= \sum_{m\bf k} \frac{e^{-\beta m\epsilon_{\bf k-}}}{Z} |\langle 1_{\bf k+} | H_{\rm R, \bf k}^{ss'}| m_{\bf k-} \rangle |^2 \\
 &\times \delta(\hbar\omega - \Delta_{\bf k}),\nonumber
\end{align}
where $\epsilon_{\bf k+}$ ($\epsilon_{\bf k-}$)is the magnon energy of the upper (lower) band, and $\Delta_{\bf k}=\epsilon_{\bf k +}-\epsilon_{\bf k-}$ is the magnon band gap. 

\begin{figure}
    \centering
    \includegraphics[width=\linewidth]{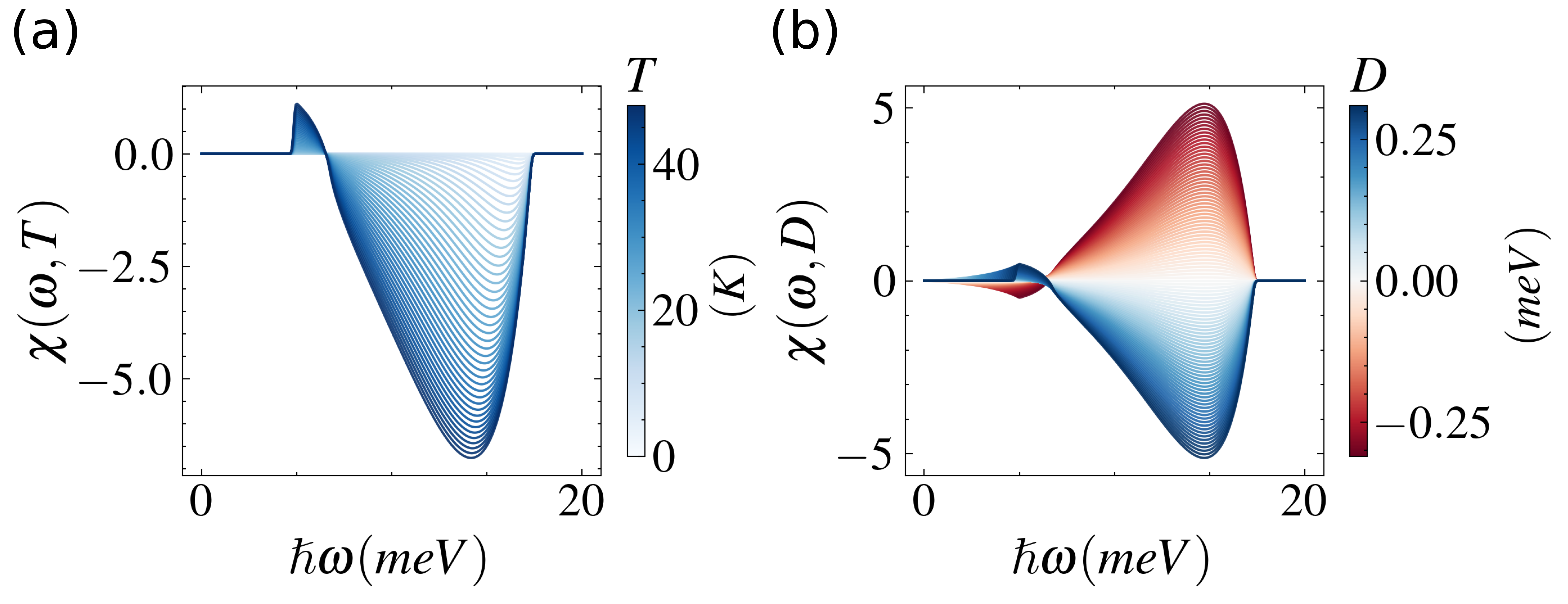}
    \caption{Thermal Raman circular dichroism as a function of frequency, for a topological monolayer of the ferromagnet CrI\textsubscript{3} for temperatures below the Curie temperature $T_c=45$~K. (a) Results for different temperatures for $D = 0.31$ meV. The positive peak indicates the topological band gap at the $K$ and $K^\prime$ points. The negative broadband signal originates from the high thermal population at the energy minimum around $\Gamma$ point. (b) Signal for different $D$ at $T = 35$~K. The signal vanishes at $D = 0$, corresponding to the topologically trivial limit. Model parameters are taken from Ref.~\cite{Chen2018}.}
    \label{fig:frequency_resolved_thermal_RCD}
\end{figure}

The finite-temperature RCD, referred as thermal RCD (TRCD) in the following,  is defined as $\chi(\omega,T) = \sum_{s'} P^{Rs'}(\omega,T) - P^{Ls'}(\omega,T)$ and can be explicitly written as
\begin{equation}
    \chi(\omega,T) = \sum_{m\bf k} K_0^2\frac{e^{-\beta m \epsilon_{\bf k-}}}{Z}\chi_{{\bf k},+-}\delta(\hbar \omega -\Delta_{\bf k}),
    \label{eq:thermal-RCD-FM}
\end{equation}
where 
\begin{equation}
\chi_{{\bf k}}=\sum_{s'} \left(\vert \langle 1_{{\bf k},+}|H^{Rs'}_{R,\bf k}|m_{{\bf k},-}\rangle\vert^2-\vert \langle 1_{{\bf k},+}|H^{Ls'}_{R,\bf k}|m_{{\bf k},-}\rangle\vert^2\right)\,.
\end{equation}
Using Eq.~\eqref{eq:thermal-RCD-FM} together with the parameters for $\text{CrI}_3$ measured in Ref.~\cite{Chen2018}, the TRCD can be immediately obtained after transforming the LMCs into the magnon basis~\cite{supplementary_material,Bostroem2023}. The model predicts a topological ferromagnetic magnon insulator for finite DMI. The TRCD is shown as a function of frequency and temperature in Fig.~\ref{fig:frequency_resolved_thermal_RCD}, remaining below the Curie temperature $T_c=45$~K for CrI\textsubscript{3}~\cite{Chen2018}. The TRCD consists of a peak dominated by the $K$ and $K^\prime$ points, and a broadband component arising from thermally weighted contributions across the Brillouin zone. The entire TRCD signal vanishes in the topologically trivial phase ($D = 0$). 

The TRCD in momentum space is related to the Berry curvature of the lower magnon band $\Omega_-({\bf k})$~\cite{Bostroem2023,Niu_2010,supplementary_material} via 
\begin{equation}
    \chi_{\bf k} = -\Omega_{-}({\bf k}) \varrho_{\bf k}  , \label{eq:RCD-Berry-curvature}
\end{equation}
where $\varrho_{\bf k}$ is a weighting function that depends on the local geometry of the energy gap (see SM~\cite{supplementary_material} for details). This result is analogous to what was found in Ref.~\cite{Bostroem2023} for a canted AFM at zero temperature.
Assuming that the local geometry of the energy gap is inversion symmetric, i.e., $\varrho_{-\bf k} = \varrho_{\bf k}$, the TRCD will always vanish when the system possesses time-reversal symmetry.

%-----------------------------------------------------------------
% Figure: Thermal RCD
%-----------------------------------------------------------------

%-----------------------------------------------------------------
% Discussion and outlook
%-----------------------------------------------------------------
\textit{Conclusion and discussion}. 
In this work, we have demonstrated that the FL vertex can be obtained via an LMC formalism, akin to electronic systems,
by taking momentum derivatives of the quadratic spin Hamiltonian $H^0_M({\bf k})$. This method is valid whenever the condition in Eq.~\eqref{eq:charge-neutral-k-A-dependence} is satisfied, and allows for the inclusion of quadratic spin interactions beyond the Heisenberg exchange, such as DMI and symmetric anisotropic exchange. Crucially, this shortcut is generally applicable to Mott insulating materials. It remains applicable to Mott-Hund insulators, provided that the Raman operator arises from direct hopping (see Ref.~\cite{Yang2021_nonFL}). Many ferromagnetic insulators, such as pyrochlore ferromagnets $\mathrm{Lu}_2\mathrm{V}_2\mathrm{O}_7$ \cite{Onose2010} and $\mathrm{CrI}_3$ \cite{Chen2018}, fall into this Mott-Hund insulator class. While assessing the magnitude of this contribution compared to ligand-mediated hopping terms is beyond the scope of this work, a detailed discussion of such "non-FL" terms in the specific case of $\beta$-$\mathrm{Li}_2\mathrm{IrO}_3$ can be found in Ref.~\cite{Yang2021_nonFL}.

We applied our method to calculate the TRCD of a topological ferromagnetic magnon insulator described by Heisenberg exchange, DMI, and easy-axis anisotropy. We found that the TRCD vanishes whenever the system is in a topologically trivial regime.
We showed that the TRCD is connected to the Berry curvature by a weighting function $\varrho_{\bf k}$ containing information on the quantum geometry of the bands. This is consistent with recent results in spin liquids showing that the RCD in momentum space is related to quantum metric and second derivatives of the eigenstates~\cite{koller2025ramancirculardichroismquantum}. 
The link between RCD and the quantum geometry of magnon bands is most explicit in the two-band limit \cite{Bostroem2023}. In multi-band systems, such as the four-band canted AFM in a honeycomb lattice, this connection becomes more intricate due to the presence of multiple scattering channels. Nevertheless, a comprehensive relationship can be established by accounting for all scattering processes~\cite{our_long_paper}. Recently, bicircular Raman spectroscopy has been proposed to probe the quantum geometry of altermagnetic magnon bands~\cite{yuan2025quantumgeometryaltermagneticmagnons}. In the broader context of electronic systems, alternative probes of quantum geometry have also been identified, including magnetic circular dichroism~\cite{ghosh_probing_2024} and quantum optical measurement schemes~\cite{Werner_2023_qo_qg_topo_inv}. We expect that a full analogy between the optical response and quantum geometry -- similar to that developed for electronic systems~\cite{Topp2021, Ahn2022} -- should extend to bosonic systems. We leave this exploration for future work.

Lastly, we address the treatment of the external electromagnetic field. The standard approach typically employs a quantized vector potential in the Peierls substitution before employing a perturbation scheme~\cite{Shastry_Shraiman_1990,ShastryShraiman1991,Yang2021_nonFL,Lee2010}, whereas we used a classical time-dependent field and quantized the field at the very end of the derivation. These two frameworks are expected to converge in the limit of weak light-matter interaction. However, this equivalence is not universal, as discussed in Ref.~\cite{Michael2020}. Investigating the discrepancies between semiclassical and fully quantum mechanical descriptions of optical observables in magnon systems remains an open and interesting question. We note that while the Peierls coupling remains the dominant mechanism for Raman spectroscopy, other light-matter couplings in a magnetic system -- such as the Aharonov-Casher effect -- could in principle be considered. However, the $({\bf k},{\bf E})$ relation is structurally trivial in these cases, so we have omitted them.

In summary, this Letter introduces a shortcut for deriving the FL vertex in  quadratic spin models, providing a practical route to evaluate the optical observables involving two-photon two-magnon processes. This work lays the foundation for a geometric understanding of the optical response in magnonic systems.

\begin{acknowledgments} 
\noindent \textit{Acknowledgments}.---
We thank Johannes Knolle and Achim Rosch for fruitful discussions. SVK and MAS acknowledge funding by the Deutsche Forschungsgemeinschaft (DFG, German Research Foundation)- 531215165 (Research Unit ‘OPTIMAL’)). YSL and SVK acknowledge funding by the DFG Project-ID 541503763 (Research Unit  ‘ChiPS’,). MAS was funded by the European Union (ERC, CAVMAT, project no. 101124492). EVB acknowledges support from the Cluster of Excellence “CUI: Advanced Imaging of Matter”–EXC 2056–project ID 390715994, SFB-925 “Light induced dynamics and control of correlated quantum systems”–project ID 170620586 of the Deutsche Forschungsgemeinschaft (DFG), and the European Research Council (ERC-2024-SyG-UnMySt–101167294). Views and opinions expressed are however those of the author(s) only and do not necessarily reflect those of the European Union or the European Research Council. Neither the European Union nor the European Research Council can be held responsible for them. 
\end{acknowledgments}

%-----------------------------------------------------------------
% Appendix
%-----------------------------------------------------------------
%\appendix

%-----------------------------------------------------------------
% Bibliography
%-----------------------------------------------------------------

\bibliography{reference,reference_famous,reference_non_hermitian, reference_other_bosonic, Mott-spin-chirality}

%%%%%%%%%% Merge with supplemental materials %%%%%%%%%%
\pagebreak
\widetext
\begin{center}
\textbf{\large Supplementary Material for "Light-Matter-Coupling formalism for magnons: probing quantum geometry with light"}
\end{center}
\renewcommand{\theequation}{S\arabic{equation}}
\renewcommand{\thefigure}{S\arabic{figure}}
\setcounter{equation}{0}
\setcounter{figure}{0}
\setcounter{secnumdepth}{3}
\setcounter{page}{1} % Reset the page counter to 1
\makeatletter
\renewcommand{\thesection}{\Roman{section}}
\renewcommand{\thesubsection}{\Alph{subsection}}
\renewcommand{\thepage}{S\arabic{page}}
\makeatother
%-----------------------------------------------------------------
% Outline of the supplementary material
%-----------------------------------------------------------------
\section{Outline}
In this Supplementary Material, we present detailed derivations that demonstrate the validity of Eq.~\eqref{eq:charge-neutral-k-A-dependence} in the main text and establish the connection between RCD and Berry curvature given by Eq.~\eqref{eq:RCD-Berry-curvature} in the main text. Specifically, in Sec.~\ref{sec:Mott-Hubbard-perturbatin-Peierls-phase}, we apply a time-dependent Schrieffer-Wolff transformation to the Mott-Hubbard model and perform a perturbation expansion in powers of $t/U$ with a time-dependent Peierls phase, following the approaches of Kitamura \textit{et al.}~\cite{Kitamura2017} and Eckstein \textit{et al.}~\cite{Eckstein2017}. In Sec.~\ref{sec:2D-FM-magnon-insulator}, we apply our formalism to calculate the Fleury Loudon (FL) Raman vertex  to a two-band ferromagnetic magnon Hamiltonian that fulfills the condition in Eq.~\eqref{eq:even-combination-of-k-eA-suppl} and derive an expression of the RCD in terms of the Berry curvature of the magnon bands in Eq.~\eqref{eq:two-band-RCD-Berry-cruvature}. The details of the model of a monolayer $\mathrm{CrI_3}$ used in the main text are shown in Sec.~\ref{sec:example-2D-FM-honeycomb-CrI3}.

%-----------------------------------------------------------------
% Perturbative expansion of the Mott-Hubbard model in orders of t/U with Peierls phase
%-----------------------------------------------------------------
\section{Perturbative expansion of the Mott-Hubbard model with Peierls phase}\label{sec:Mott-Hubbard-perturbatin-Peierls-phase}
In this section, we provide a derivation that demonstrates the validity of the condition
\begin{equation}
    H({\bf k},e{\bf A})= \frac{1}{2}\squarebracket{H({\bf k}-e{\bf A})+H({\bf k}+e{\bf A})}\, ,  \label{eq:even-combination-of-k-eA-suppl}
\end{equation}
up to the second order in the perturbation expansion of $t/U$ for a spin Hamiltonian coupled to light. Furthermore, we show that this condition generally breaks down at higher orders. We begin with a single-band Hubbard model (Eq.~\eqref{eq:micro-Hubbard-Mott-model} in the main text),
\begin{equation}
    H_{\text{micr}} = -\sum_{\langle ij\rangle \sigma} t_{ij} \left(c_{i\sigma}^\dagger c_{j\sigma}+\text{h.c.} \right) + U\sum_i n_{i\uparrow} n_{i\downarrow}\,,
\end{equation}
in the Mott limit $0<t_{ij}\ll U$, where $t_{ij}\in \mathbb{R}$ is the hopping amplitude between site $i$ and $j$, and $U$ is the on-site Coulomb potential. In the presence of an external electromagnetic field, a standard approach to treat the coupling is to introduce a Peierls phase, 
\begin{equation}
    \widehat{{H}}(t)=  -\sum_{\langle ij\rangle \sigma} \left(t_{ij}e^{-ie\int_i^j{\bf A}({\bf r},t)\cdot d{\bf r}}c_{i\sigma}^\dagger c_{j\sigma}+\text{h.c.}\right) + U\sum_i n_{i\uparrow} n_{i\downarrow}.
\end{equation}
Here, we have set $\hbar=1$. Let $\theta_{ij}(t)=-e\int_i^j {\bf A}({\bf r},t)\cdot d{\bf r}$. Applying the dipole approximation, that is, the vector potential is slowly varying between two sites $i$ and $j$, we can re-write $\theta_{ij}(t)$ as $-e{\bf A}(t)\cdot {\bf d}_{ij}$ and define an effective hopping amplitude $\tilde{t}_{ij}(t)\coloneqq  t_{ij}e^{i\theta_{ij}(t)}\in \mathbb{C}$. Following the Floquet-Magnus expansion used in Ref.~\cite{Kitamura2017}, we first re-write the time-dependent Schrieffer-Wolff (SW) transformation $\widetilde{H}(t)=e^{i\widehat{S}(t)}(\widehat{{H}}(t)-i\partial_t)e^{-i\widehat{{S}}(t)}$ into a recursive relation with a unitary operator $e^{i\widehat{{S}}(t)}$
\begin{equation}
    \widetilde{{H}}(t) +\frac{\partial \widehat{{S}}}{\partial t} = \widehat{{H}}(t) - \sum_{n=1}^\infty \frac{B_n}{n!}\text{ad}^n_{i\widehat{{S}}(t)} 
    \left((-1)^{n-1}\widehat{{H}}(t)+\widetilde{H}(t) \right),
\end{equation}
where $B_n$ is the Bernoulli's number, $\widehat{{S}}(t)$ is the SW transformation generator and $\text{ad}_{i\widehat{{S}}(t)}\cdot=\left[i\widehat{{S}}(t), \cdot\right]$. In the Mott limit $|\tilde{t}_{ij}|=|t_{ij}|\ll U$, we can split the Hamiltonian $\widehat{{H}}(t)=-\lambda\widehat{T}(t)+\widehat{{U}}(t)$ into a hopping term $\widehat{{T}}(t) = \sum_{ij\sigma} \left(\tilde{t}_{ij} c_{i\sigma}^\dagger c_{j\sigma}+\text{h.c}\right)$ and an on-site potential term $\widehat{U}=U\sum_i n_{i\uparrow}n_{i\downarrow}$, where $\lambda$ is a parameter to keep track of the order of expansion. We relax the constraint of nearest-neighbor hopping to study the general contribution from the hopping term.

Analogously for the transformed Hamiltonian $\widetilde{{H}}(t) =\tilde{H}(t)+\widehat{{U}}(t)$, we obtain Eq.~(5) in  Ref.~\cite{Kitamura2017}
\begin{equation}
    \tilde{H}(t)+\left[\widehat{{U}},i\widehat{{S}}(t)\right]+\partial_t\widehat{{S}}(t) = -\lambda \widehat{T}(t) -\sum_{n=1}^\infty  \frac{B_n}{n!}\text{ad}^n_{i\widehat{{S}}(t)}\left[(-1)^n\lambda \widehat{{T}}(t)+\tilde{H}(t)\right].
\end{equation}

By expanding $\tilde{H}(t)$ and $\widehat{{S}}(t)$ in order of $\lambda$ to keep track of the perturbation order,
\begin{equation}
    \tilde{H}(t) = \sum_{n=1}^\infty \lambda^n \tilde{H}^{(n)}(t),\quad \widehat{{S}}(t)=\sum_{n=1}^\infty \lambda^n \widehat{S}^{(n)}(t),
\end{equation}
and defining $\widehat{M}^{(n)}=\tilde{H}^{(n)}+\left[\widehat{U},i\widehat{{S}}^{(n)}(t)\right]+\partial_t\widehat{{S}}^{(n)}(t)$, we have $\widehat{M}^{(n)}$ up to fourth order (with $\lambda =1$)
\begin{align}
    \widehat{M}^{(1)} &=-\widehat{T}(t)  \label{eq:first-order-TDSWT} \\
	\widehat{M}^{(2)} &= -\frac{1}{2} \left[i\widehat{S}^{(1)}(t),\widehat{T}(t)-\tilde{H}^{(1)}(t)\right] \label{eq:second-order-TDSWT}, \\
	\widehat{M}^{(3)} &= -\frac{1}{2} 
    \left[i\widehat{S}^{(2)}(t),\widehat{T}(t)-\tilde{H}^{(1)}(t)\right] +\frac{1}{2}\left[i\widehat{S}^{(1)}(t),\tilde{H}^{(2)}(t) \right]\nonumber \\ &\quad -\frac{1}{12}\left[i\widehat{S}^{(1)}(t),\left[i\widehat{S}^{(1)}(t),\widehat{T}(t)+\tilde{H}^{(1)}(t)  \right]\right]\label{eq:third-order-TDSWT} , \\
	\widehat{M}^{(4)} &= -\frac{1}{2}\left[i\widehat{S}^{(3)}(t),\widehat{T}(t)-\tilde{H}^{(1)}(t)\right] +\frac{1}{2}\left[i\widehat{S}^{(2)}(t),\tilde{H}^{(2)}(t)\right] + \frac{1}{2}\left[i\widehat{S}^{(1)}(t),\tilde{H}^{(3)}(t)\right]\nonumber \\ &\quad -\frac{1}{12}\left[i\widehat{S}^{(1)}(t),\left[i\widehat{S}^{(1)}(t),\tilde{H}^{(2)}(t)\right]\right]-\frac{1}{12}\left[i\widehat{S}^{(1)}(t),\left[i\widehat{S}^{(2)}(t),\widehat{T}+\tilde{H}^{(1)}(t)\right]\right]  \nonumber \\ &\quad -\frac{1}{12}\left[i\widehat{S}^{(2)}(t),\left[i\widehat{S}^{(1)}(t),\widehat{T}+\tilde{H}^{(1)}(t)\right]\right] \label{eq:fourth-order-TDSWT}. 
\end{align}
These equations are Eqs.~(6-9) of Ref.~\cite{Kitamura2017}. Since the procedure of the SW transformation consists in removing the off-diagonal terms present in $\widehat{{S}}^{(n)}(t)$ at order $n$, we have to identify what terms are off-diagonal, that is, coupling between low-energy and high-energy subspaces. At half-filling, the low-energy subspace is the space of configurations with only a single electron occupying each site, while the high-energy subspace is the space of configurations with any double occupancy. Any hopping term inducing a double occupancy is thus off-diagonal. Therefore, $\widehat{{S}}^{(n)}(t)$ can be decomposed into $\widehat{{S}}^{(n)}_{-1}(t)+\widehat{{S}}_{+1}^{(n)}(t)$ (diagonal terms can be absorbed in $\tilde{H}^{(n)}(t)$); while the hopping operator corresponding to a single electron hopping can be written as
\begin{equation}
    \widehat{{T}}(t) =  \widehat{{T}}_{-1}(t)+ \widehat{{T}}_0(t) +  \widehat{{T}}_{+1}(t).
\end{equation}
Here, the notation $\pm 1$  indicates the addition or removal of double occupancy by the operator. For the subindex $0$, the operator does not add or remove any double occupancy.

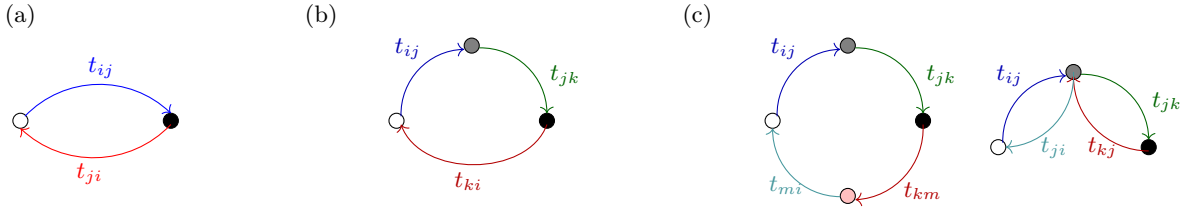
\begin{figure}[!ht]
	\centering
	\begin{tikzpicture}
	[circle A/.style={radius=1mm},
	circle B/.style={fill=black, radius=1mm},
	circle C/.style={fill=gray, radius=1mm},
	circle D/.style={fill=pink, radius=1mm}]
		\begin{scope}[]
			\draw[circle A] (-1,0) circle;
			\draw[circle B] (1, 0) circle;
			\draw[->, blue] (-0.94,0.06) to [bend left=45] (1,0.1) node[above left, xshift=-18, yshift=10]{$t_{ij}$};
			\draw[->, red] (1,-0.05) to [bend left=45] (-1, -0.1)node[below right, xshift=18, yshift=-10]{$t_{ji}$};
			\node[label= (a)] at (-1,1) {};
		\end{scope}
		
		\begin{scope}[xshift=5cm]
			\draw[circle A] (-1,0) circle;
			\draw[circle B] (1, 0) circle;
			\draw[circle C] (0, 1) circle;
			\draw[->, blue!70!black] (-0.94,0.06) to [bend left=45] (-0.1,0.95) node[above left, xshift=-13, yshift=-7]{$t_{ij}$};
			\draw[->, green!40!black] (0.1,0.97) to [bend left=45] (1,0.1) node[above right, xshift=-2, yshift=7]{$t_{jk}$};
			\draw[->, red!70!black] (1,-0.05) to [bend left=70] (-0.94, -0.05)node[below right, xshift=17, yshift=-15]{$t_{ki}$};
			\node[label= (b)] at (-2,1) {};
		\end{scope}
		
		\begin{scope}[xshift=10cm]
			\draw[circle A] (-1,0) circle;
			\draw[circle B] (1, 0) circle;
			\draw[circle C] (0, 1) circle;
			\draw[circle D] (0, -1) circle;
			\draw[->, blue!70!black] (-0.94,0.06) to [bend left=45] (-0.1,0.95) node[above left, xshift=-13, yshift=-7]{$t_{ij}$};
			\draw[->, green!40!black] (0.1,0.97) to [bend left=45] (1,0.1) node[above right, xshift=-2, yshift=7]{$t_{jk}$};
			\draw[->, red!70!black] (1,-0.05) to [bend left=45] (0.1, -1.05)node[below right, xshift=15, yshift=10]{$t_{km}$};
			\draw[->, Aquamarine!70!black] (-0.1, -1.01) to [bend left=45] (-1, -0.1)node[below left, xshift=15, yshift=-15]{$t_{mi}$};
			\node[label= (c)] at (-2,1) {};
		\end{scope}
		\begin{scope}[xshift=13cm, yshift=-10]
			\draw[circle A] (-1,0) circle;
			\draw[circle B] (1, 0) circle;
			\draw[circle C] (0, 1) circle;
			\draw[->, blue!70!black] (-0.94,0.06) to [bend left=45] (-0.1,0.95) node[above left, xshift=-13, yshift=-7]{$t_{ij}$};
			\draw[->, green!40!black] (0.1,0.97) to [bend left=45] (1,0.1) node[above right, xshift=-2, yshift=7]{$t_{jk}$};
			\draw[->, red!70!black] (1,-0.05) to [bend left=45] (0, 0.95)node[below left, xshift=20, yshift=-20]{$t_{kj}$};
			\draw[->, Aquamarine!70!black] (0, 1.) to [bend left=45] (-0.9, -0)node[below right, xshift=10, yshift=7]{$t_{ji}$};
		\end{scope}
	\end{tikzpicture}
	\caption{Possible hopping paths for second, third, and fourth order perturbation of Hubbard-Mott model at half-filling. }
	\label{fig:hopping-path}
\end{figure}

Further, at half-filling, we expect the following commutation relation
\begin{equation}
    \left[\widehat{U},\widehat{{S}}^{(n)}_{\pm 1}\right]= \pm U\widehat{{S}}^{(n)}_{\pm 1}
\end{equation}
holds. Before going into the details of expansion, we note a general fact: since the effective Hamiltonian $\tilde{H}(t)$ describes the effective picture of the original system in the low-energy subspace, at all orders the expansion must start with creating a double occupancy $\widehat{T}_{+1}(t)$ and end with destroying a double occupancy $\widehat{{T}}_{-1}(t)$. See Fig.~\ref{fig:hopping-path} for possible virtual hopping paths at second, third, and fourth orders.
%-----------------------------------------------------------------
% Figure: Possible hoping paths
%-----------------------------------------------------------------

\subsection{First order expansion}
To first order, the corresponding equation is Eq.~\eqref{eq:first-order-TDSWT}, which gives
\begin{equation}
    \tilde{H}^{(1)}(t) + \left[\widehat{{U}},i\widehat{S}^{(1)}_{+1}(t)\right]+\left[\widehat{{U}},i\widehat{S}^{(1)}_{-1}(t)\right]+\partial_t \widehat{{S}}^{(1)}_{+1}(t)+\partial_t \widehat{{S}}^{(1)}_{-1}(t) = -\widehat{{T}}_{+1}(t)-\widehat{{T}}_{-1}(t) -\widehat{{T}}_0(t).
\end{equation}
By matching the number of double occupancy of each term, we have three separate equations
\begin{align}
    \tilde{H}^{(1)}(t) &= -\widehat{{T}}_0(t), \\
    \left[\partial_t\pm iU\right]\widehat{{S}}_{\pm 1}^{(1)}(t) &= -\widehat{T}_{\pm 1}(t). \label{eq:SW-operator-first-order} 
\end{align}
By restricting these to the low-energy subspace via the projection operator $\widehat{P}=\prod_i \left({\bf 1}-n_{i\uparrow}n_{i\downarrow}\right)$, we have $\widehat{{P}}\tilde{H}^{(1)}(t)\widehat{{P}}=0$ because no single electron hopping in the low-energy subspace is possible without creating a double occupancy. To obtain a solution that is well-defined up to time $t$, we use the retarded Green's function  corresponding to the differential operators $\partial_t\pm iU$ as
\begin{equation}
    G^R(t,t') = -ie^{-\eta (t-t')}e^{-idU(t-t')} \theta(t-t')= -ie^{\eta t'}e^{-idU(t-t')} \theta(t-t').
\end{equation}
with $\eta\rightarrow 0^+$ and $d=\pm 1$. The convergence factor $e^{\eta t'}$ is to ensure that the solution vanishes at $t'\rightarrow -\infty$. The second equality is obtained because we consider only finite time $t$. Then the SW generator to first order is obtained as the convolution of retarded Green's function and the function on the right hand side of Eq.~\eqref{eq:first-order-TDSWT}, that is,
\begin{equation}
    \widehat{{S}}^{(1)}_{\pm 1}(t) = -i \int_{-\infty}^{+\infty} G^R(t,t') \widehat{T}_{\pm1}(t') dt' = -\int_{-\infty}^{+\infty} \theta(t-t')e^{\eta t'} e^{\mp iU(t-t')} \widehat{{T}}_{\pm 1}(t') dt'. \label{eq:SW-operator-first-order-2}
\end{equation}
As a result of the low-energy subspace projection, the effective Hamiltonian in the low-energy subspace always starts at second order
\begin{equation}
    \tilde{H}(t) = \sum_{n=2} \tilde{H}^{(n)}(t).
\end{equation}
\subsection{Second order expansion}
To second order, the corresponding equation is Eq.~\eqref{eq:second-order-TDSWT}. The procedure is exactly the same as for the first order calculation, by matching the number of occupancy between the two sides. We then obtain three separate equations as
\begin{align}
    \tilde{H}^{(2)}(t) &= -\frac{i}{2}\left[\widehat{S}_{-1}^{(1)}(t),\widehat{T}_{+1}(t)\right]-\frac{i}{2}\left[\widehat{S}^{(1)}_{+1}(t),\widehat{T}_{-1}(t)\right], \label{eq:quadratic-effective-TDSWT-Hamiltonian} \\
	\partial_t \widehat{S}^{(2)}_{\pm 1}(t) \pm  iU\widehat{S}^{(2)}_{\pm 1}(t) &= -i\left[\widehat{S}^{(1)}_{\pm 1}(t),\widehat{T}_0(t)\right]. \label{eq:time-dependent-SW-transformation-second-order-expansion}
\end{align}
Using Eq.~\eqref{eq:SW-operator-first-order-2}, the above equations read
\begin{align}
    \tilde{H}^{(2)}(t) &= \frac{i}{2}\int_{-\infty}^{+\infty} \theta(t-t') e^{\eta t'}\left(e^{iU(t-t')}\left[\widehat{T}_{-1}(t'),\widehat{{T}}_{+1}(t)\right]+ e^{-iU(t-t')}\left[\widehat{T}_{+1}(t'),\widehat{T}_{-1}(t)\right] \right) dt' \\ 
    \widehat{S}^{(2)}_{\pm}(t) &= i\int_{-\infty}^{+\infty}\int_{-\infty}^{+\infty} \theta(t-t')\theta(t'-t'') e^{\eta(t'+t'')}e^{\mp i U(t-t')} e^{\mp iU(t'-t'')}[\widehat{T}_{\pm 1}(t''),\widehat{T}_0(t')] dt'' dt'. \label{eq:SW-operator-second-order}
\end{align}
To get a closed form of the Hamiltonian in Eq.~\eqref{eq:time-dependent-SW-transformation-second-order-expansion}, we first re-write the hopping term as
\begin{equation}
    \widehat{T}(t) = \sum_{ij}\sum_{\sigma\sigma'} c^\dagger_{i\sigma} M^{ij}_{\sigma\sigma'} c_{j\sigma'} e^{i\theta_{ij}(t)} +\text{h.c}, \quad M^{ij}_{\sigma\sigma'}= (-t_{ij}{\bf 1})_{\sigma\sigma'}.
\end{equation}
Then, we have
\begin{align}
    \widehat{T}_{-1}(t')\widehat{T}_{+1}(t) &=\sum_{ ij}  \sum_{\sigma_1\sigma_2\sigma_3\sigma_4} c_{i\sigma_1}^\dagger M^{ij}_{\sigma_1\sigma_2}c_{j\sigma_2} e^{i\theta_{ij}(t')} c_{j\sigma_3}^\dagger M^{ij}_{\sigma_3\sigma_4}c_{i\sigma_4} e^{i\theta_{ji}(t)} +\text{h.c.} \nonumber \\
    &= 2\sum_{ij}\text{tr}\squarebracket{M^{ij}M^{ji}\roundbracket{\frac{1}{2}+{\bf S}_i\cdot \boldsymbol{\tau}} \cos(\theta_{ij}(t')-\theta_{ij}(t))} \nonumber \\
	& + 2\sum_{ij}\text{tr}\squarebracket{M^{ij}\roundbracket{\frac{1}{2}+{\bf S}_j\cdot \boldsymbol{\tau}}M^{ji}\roundbracket{\frac{1}{2}+{\bf S}_i\cdot \boldsymbol{\tau}} \cos(\theta_{ij}(t')-\theta_{ij}(t))},
\end{align}
where we have applied the Jordan map $c_{i\sigma}^{\dagger}c_{i\sigma'}=\left(\frac{1}{2}n_i+\mathbf{S}_{i}\cdot\boldsymbol{\tau}\right)_{\sigma'\sigma} = \left(\frac{1}{2}{\bf 1}+\mathbf{S}_{i}\cdot\boldsymbol{\tau}\right)_{\sigma'\sigma}$ where in the second equality we have re-written the sum in the spin index into a trace.
%projected the map to the low-energy subspace. 

After the low-energy subspace projection, we obtain
\begin{equation}
    \tilde{H}^{(2)}(t) = \sum_{ij} \text{Im} \int_{-\infty}^{+\infty} \theta(t-t') e^{\eta t'} e^{iU(t-t')} \widehat{T}_{-1}(t')\widehat{T}_{+1}(t) dt'.
\end{equation}
After some steps for computing the trace, we obtain the second order effective spin Hamiltonian
\begin{equation}
    \tilde{H}^{(2)}(t) =\sum_{ij} J_{ij}(t) {\bf S}_i\cdot {\bf S}_j
\end{equation}
with the modified Heisenberg coefficient $J_{ij}(t)$ as
\begin{equation}
    J_{ij}(t)  = 4|t_{ij}|^2 I_{ij}(t), \quad I_{ij}(t)=\text{Im} \int_{-\infty}^{+\infty} \theta(t-t') e^{\eta t'}e^{iU(t-t')}\cos(\theta_{ij}(t')-\theta_{ij}(t)) dt'. 
\end{equation}
This is the result derived in Ref.~\cite{Eckstein2017}. If we compute the integral and take the limit ${\bf A}\rightarrow {\bf 0}$, this Hamiltonian recovers the standard second order SW transformation of the single band Hubbard-Mott model with coefficient $4t^2/U$ when $t_{ij}=t$ is uniform. After a Fourier transform, for a bipartite sublattice the pair-wise spin terms become $\sum_{ij}{\bf S}_{A,i}\cdot {\bf S}_{B,j}=\sum_{\bf k}\sum_{{\bf d}_{ij}}{\bf S}_{A,{\bf k}}\cdot {\bf S}_{B,-{\bf k}}e^{-i{\bf k}\cdot {\bf d}_{ij}}$ (if $t_{ij}=t$ for all sites $i$ and $j$), where $A$,$B$  label the sublattices. 
The cosine factor can be recast into an exponential form, which gives a $({\bf k}-e{\bf A})$ and $({\bf k}+e{\bf A})$ dependence. Hence, the condition for the shortcut Eq.~\eqref{eq:even-combination-of-k-eA-suppl} holds for the low-energy effective spin Hamiltonian to second order in $t/U$. We are then allowed to replace the vector potential derivatives by ${\bf k}$-derivatives, and the Fleury-Loudon vertex is recovered. Here, we considered only electron hopping $-t_{ij}\delta_{\sigma\sigma'}$ in the tensor $M^{ij}_{\sigma\sigma'}$ for simplicity. 
However, one can see that the derivation holds for a general $M^{ij}_{\sigma\sigma'}$. For example, we can add a Rashba spin-orbit term $M^{ij}_{\sigma\sigma'}=(-t_{ij}{\bf 1}+\boldsymbol{\alpha}_{ij}\cdot \boldsymbol{\tau})_{\sigma\sigma'}$ with Rashba coupling vector $\boldsymbol{\alpha}_{ij}$. 
This gives the Dzyaloshinskii–Moriya interaction term and symmetric anistropic interaction term with the same time-integral correction $I_{ij}(t)$ \cite{our_long_paper}. Hence, a spin model consisting only terms obtainable from $M_{\sigma\sigma'}^{ij}$ are deemed to satisfy the condition Eq.~\eqref{eq:even-combination-of-k-eA-suppl} for the shortcut, where terms which play no role in the (${\bf k},e{\bf A}$) dependence are omitted.

\subsection{Third order expansion}
To the third order, following the same procedure,
\begin{align}
    \tilde{H}^{(3)}(t) &= -\frac{i}{2} \left[\widehat{S}^{(2)}_{-1}(t),\widehat{T}_{+1}(t) \right] - \frac{i}{2}\left[\widehat{S}^{(2)}_{+1}(t),\widehat{{T}}_{-1}(t)\right] \\
    \partial_t \widehat{S}^{(3)}_{\pm 1}(t) \pm  iU\widehat{S}^{(3)}_{\pm 1}(t) &= -\left[i\widehat{S}^{(2)}_{\pm 1}(t),\widehat{T}_{0}(t)\right] + \frac{1}{2}\left[i\widehat{S}^{(1)}_{\pm 1}(t), \tilde{H}^{(2)}(t)\right]- \frac{1}{12}\left[i\widehat{S}^{(1)}_{\mp 1}(t),\left[i\widehat{S}^{(1)}_{\pm 1}(t), \widehat{T}_{\pm 1}(t)\right]\right]  \nonumber\\
	&-\frac{1}{12} \left[i\widehat{S}^{(1)}_{\pm 1}(t),\left[i\widehat{S}^{(1)}_{\pm 1}(t), \widehat{T}_{\mp 1}(t)\right]\right] -\frac{1}{12}\left[i\widehat{S}^{(1)}_{\pm 1}(t),\left[i\widehat{S}^{(1)}_{\mp 1}(t), \widehat{T}_{\pm 1}(t)\right]\right].
\end{align}
By substituting the second order SW generator $\widehat{S}^{(2)}(t)$ and second order low-energy effective Hamiltonian, we obtain
\begin{align}
    \tilde{H}^{(3)}(t) &= \frac{1}{2}\int_{-\infty}^{+\infty}\int_{-\infty}^{+\infty}\theta(t-t')\theta(t'-t'')e^{\eta(t'+t'')} e^{+iU(t-t')}e^{+iU(t'-t'')} \left[\left[\widehat{T}_{- 1}(t''),\widehat{T}_0(t')\right], \widehat{T}_{+1}(t)\right] dt'' dt' \nonumber \\
	&+ \frac{1}{2}\int_{-\infty}^{+\infty}\int_{-\infty}^{+\infty}\theta(t-t')\theta(t'-t'')e^{\eta(t'+t'')} e^{-iU(t-t')}e^{-iU(t'-t'')}\left[\left[\widehat{T}_{+1}(t''),\widehat{T}_0(t')\right], \widehat{T}_{-1}(t)\right] dt''dt', \label{eq:third-order-SW-hamiltonian}\\
    \widehat{S}^{(3)}_{\pm 1}(t)&=  \iiint  dt_3 dt_2dt_1e^{\eta(t_1+t_2+t_3)} \bigg\{ \theta(t-t_1)\theta(t_1-t_2)\theta(t_2-t_3)e^{-iU(t-t_3)} \left[\left[\widehat{T}_{\pm 1}(t_3),\widehat{T}_0(t_2)\right],\widehat{T}_0(t_1)\right] \nonumber \\
	&-\frac{1}{3}\theta(t-t_1)\theta(t_1-t_2)\theta(t_1-t_3) e^{-iU(t-t_2)}e^{-iU(t_1-t_3)}\left[\widehat{T}_{\pm 1}(t_2),\left[ \widehat{T}_{\pm 1}(t_3),\widehat{T}_{\mp 1}(t_1)\right]\right] \nonumber \\
	& -\frac{1}{3}\theta(t-t_1)\theta(t_1-t_2)\theta(t_1-t_3) e^{-iU(t-t_2)}e^{-iU(t_1-t_3)}\left[\widehat{T}_{\pm 1}(t_2),\left[ \widehat{T}_{\mp 1}(t_3),\widehat{T}_{\pm 1}(t_1)\right]\right]  \bigg\}.
\end{align}
After applying the low-energy subspace projection on Eq.~\eqref{eq:third-order-SW-hamiltonian}, the commutators of hopping terms reduce to
\begin{align}
	\left[\left[\widehat{T}_{- 1}(t''),\widehat{T}_0(t')\right], \widehat{T}_{+1}(t)\right] &\rightarrow \widehat{T}_{- 1}(t'')\widehat{T}_{0}(t')\widehat{T}_{+1}(t), \\
	 \left[\left[\widehat{T}_{+1}(t''),\widehat{T}_0(t')\right], \widehat{T}_{-1}(t)\right] &\rightarrow -\widehat{T}_{-1}(t)\widehat{T}_0(t')\widehat{T}_{+1}(t'').
\end{align}
These products of hopping terms can be treated as in the previous section
\begin{align}
    \widehat{T}_{- 1}(t'')\widehat{T}_{0}(t')\widehat{T}_{1}(t) &= \sum_{ijk} \sum_{\sigma_1\sigma_2\sigma_3\sigma_4\sigma_5\sigma_6} c_{i\sigma_1}^\dagger M^{ik}_{\sigma_1\sigma_2} e^{i\theta_{ik}(t'')}c_{k\sigma_2} c_{k\sigma_3}^\dagger M^{kj}_{\sigma_3\sigma_4}e^{i\theta_{kj}(t')} c_{j\sigma_4} c_{j\sigma_5}^\dagger M^{ji}_{\sigma_5\sigma_6} e^{i\theta_{ji}(t)}c_{i\sigma_6}  +\text{h.c.} \nonumber \\
    &= \sum_{ijk} \bigg\{ \text{tr}\squarebracket{M^{ik}M^{kj}M^{ji}\roundbracket{\frac{1}{2}+{\bf S}_i\cdot \boldsymbol{\tau}} }  - \text{tr}\squarebracket{M^{ik}M^{kj}\roundbracket{\frac{1}{2}+{\bf S}_j\cdot \boldsymbol{\tau}} M^{ji}\roundbracket{\frac{1}{2}+{\bf S}_i\cdot \boldsymbol{\tau}} }  \nonumber \\
	&- \text{tr}\squarebracket{M^{ik} \roundbracket{\frac{1}{2}+{\bf S}_k\cdot \boldsymbol{\tau}}  M^{kj}M^{ji}\roundbracket{\frac{1}{2}+{\bf S}_i\cdot \boldsymbol{\tau}} } \nonumber \\
	& + \text{tr}\squarebracket{M^{ik} \roundbracket{\frac{1}{2}+{\bf S}_k\cdot \boldsymbol{\tau}}  M^{kj}\roundbracket{\frac{1}{2}+{\bf S}_j\cdot \boldsymbol{\tau}} M^{ji}\roundbracket{\frac{1}{2}+{\bf S}_i\cdot \boldsymbol{\tau}} } \bigg\} e^{i(\theta_{ik}(t'')+\theta_{kj}(t')+\theta_{ji}(t))}  +\text{h.c.}  \nonumber  \\
    &= -\sum_{ijk} \left[2\text{Re}(\tilde{t}_{ik}\tilde{t}_{kj}\tilde{t}_{ji}) ({\bf S}_i\cdot {\bf S}_k+{\bf S}_j\cdot {\bf S}_i+-{\bf S}_k\cdot {\bf S}_j) + 4\text{Im}(\tilde{t}_{ik}\tilde{t}_{kj}\tilde{t}_{ji}) ({\bf S}_k \times {\bf S}_j)\cdot {\bf S}_i\right] .
\end{align}
Eventually, the third order low-energy effective Hamiltonian reads
\begin{align}
    \tilde{H}^{(3)}(t) &= \sum_{ijk} I_{ijk}^{(3)}(t)	B_{ijk} + \sum_{ijk} J_{ijk}^{(3)}(t) \chi_{kji},
\end{align}
where we have defined $B_{ijk}= {\bf S}_i\cdot {\bf S}_k+{\bf S}_j\cdot {\bf S}_i-{\bf S}_k\cdot {\bf S}_j$ and the scalar spin chirality $\chi_{kji}= ({\bf S}_k \times {\bf S}_j)\cdot {\bf S}_i$ with the time-dependent coefficients
\begin{align}
    I_{ijk}^{(3)}(t) &=  -\text{Re}\int_{-\infty}^{+\infty}\int_{-\infty}^{+\infty}\theta(t-t')\theta(t'-t'') e^{\eta (t'+t'')} e^{+iU(t-t'')}t_{ik}t_{kj}t_{ji} \nonumber \\
	&\quad \times \left[ \cos(\theta_{ik}(t'')+\theta_{kj}(t')+\theta_{ji}(t))+\cos(\theta_{ik}(t)+\theta_{kj}(t')+\theta_{ji}(t''))\right] dt'' dt',\\
	J_{ijk}^{(3)}(t) &=  -2\text{Re}\int_{-\infty}^{+\infty}\int_{-\infty}^{+\infty}\theta(t-t')\theta(t'-t'') e^{\eta (t'+t'')} e^{+iU(t-t'')}t_{ik}t_{kj}t_{ji} \nonumber \\
	&\quad \times \left[ \sin(\theta_{ik}(t'')+\theta_{kj}(t')+\theta_{ji}(t))+\sin(\theta_{ik}(t)+\theta_{kj}(t')+\theta_{ji}(t'')) \right] dt'' dt'.
\end{align}
In the static field limit and undoing the dipole approximation, a flux term emerges from close loop virtual hopping  $I_{ijk}^{(3)}(t)\rightarrow 2 t_{ik}t_{kj}t_{ji}\cos (2\pi \Phi_B/\Phi_0)/U^2$ and $J_{ijk}^{(3)}(t) \rightarrow 4t_{ik}t_{kj}t_{ji}\sin(2\pi \Phi_B/\Phi_0)/U^2$, which agrees with the literature, see e.g. Refs.~\cite{Diptiman_Chitra_spin_chirality_1995, Montrunich_spin_chirality_2006, Kitamura2017}. These originate from an external field coupled to the magnetic moment created in a closed loop virtual hopping, and are incompatible with a dependence of the type $({\bf k}-e{\bf A})$ and $({\bf k}+e{\bf A})$ in the Hamiltonian. 
If we first make a dipole approximation, which is equivalent to treating zero magnetic flux $\Phi_B=0$, we still have a finite correction $2t_{ik}t_{kj}t_{ji}/U^2$ to the Heisenberg exchange coefficient~\cite{Diptiman_Chitra_spin_chirality_1995, Montrunich_spin_chirality_2006, Kitamura2017}. Moreover, the spin interaction terms which emerge with the imaginary part of the product of effective hopping amplitudes $\tilde{t}_{ij}(t)$ are so-called non-Fleury-Loudon terms, which are absent if we set the external field to zero. A dependence of the type $({\bf k}-e{\bf A})$ and $({\bf k}+e{\bf A})$ can only be recovered for Fleury-Loudon terms in third order considering exclusively virtual hopping on a straight line and between sites with the same magnitude $|{\bf d}_{ij}|=\dots=|{\bf d}_{jk}|$. This kind of virtual hopping is unavailable for odd order expansion. Therefore, the dependence of $({\bf k}-e{\bf A})$ and $({\bf k}+e{\bf A})$ in a spin Hamiltonian already does not hold at the level of the third order in the expansion in $t/U$.

\subsection{Fourth order expansion}
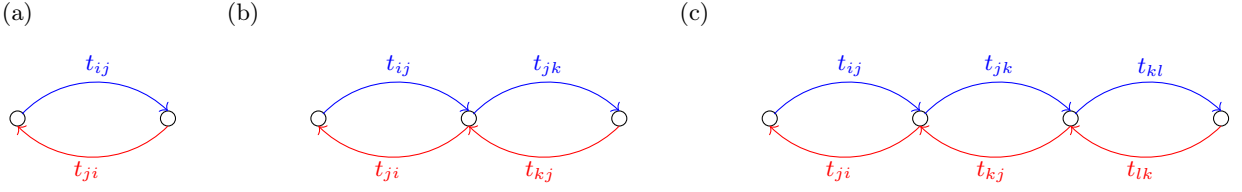
\begin{figure}[!ht]
	\centering
	\begin{tikzpicture}
	[circle A/.style={radius=1mm},
	circle B/.style={fill=black, radius=1mm},
	circle C/.style={fill=gray, radius=1mm},
	circle D/.style={fill=pink, radius=1mm}]
		\begin{scope}[]
			\draw[circle A] (-1,0) circle;
			\draw[circle A] (1, 0) circle;
			\draw[->, blue] (-0.94,0.06) to [bend left=45] (1,0.1) node[above left, xshift=-18, yshift=10]{$t_{ij}$};
			\draw[->, red] (1,-0.1) to [bend left=45] (-1, -0.1)node[below right, xshift=18, yshift=-10]{$t_{ji}$};
			\node[label= (a)] at (-1,1) {};
		\end{scope}
		
		\begin{scope}[xshift=4cm]
			\draw[circle A] (-1,0) circle;
			\draw[circle A] (1, 0) circle;
			\draw[->, blue] (-0.94,0.06) to [bend left=45] (1,0.1) node[above left, xshift=-18, yshift=10]{$t_{ij}$};
			\draw[->, red] (1,-0.1) to [bend left=45] (-1, -0.1)node[below right, xshift=18, yshift=-10]{$t_{ji}$};
			
			\draw[circle A] (3,0) circle;
			\draw[->, blue] (2+-0.94,0.06) to [bend left=45] (2+1,0.1) node[above left, xshift=-18, yshift=10]{$t_{jk}$};
			\draw[->, red] (3,-0.1) to [bend left=45] (1, -0.1)node[below right, xshift=18, yshift=-10]{$t_{kj}$};
			\node[label= (b)] at (-2,1) {};
		\end{scope}
		
		\begin{scope}[xshift=10cm]
			
		\end{scope}
		\begin{scope}[xshift=10cm, ]
			\draw[circle A] (-1,0) circle;
			\draw[circle A] (1, 0) circle;
			\draw[->, blue] (-0.94,0.06) to [bend left=45] (1,0.1) node[above left, xshift=-18, yshift=10]{$t_{ij}$};
			\draw[->, red] (1,-0.1) to [bend left=45] (-1, -0.1)node[below right, xshift=18, yshift=-10]{$t_{ji}$};
			
			\draw[circle A] (3,0) circle;
			\draw[->, blue] (2+-0.94,0.06) to [bend left=45] (2+1,0.1) node[above left, xshift=-18, yshift=10]{$t_{jk}$};
			\draw[->, red] (3,-0.1) to [bend left=45] (1, -0.1)node[below right, xshift=18, yshift=-10]{$t_{kj}$};
			
			\draw[circle A] (5,0) circle;
			\draw[->, blue] (4+-0.94,0.06) to [bend left=45] (4+1,0.1) node[above left, xshift=-18, yshift=10]{$t_{kl}$};
			\draw[->, red] (5,-0.1) to [bend left=45] (3, -0.1)node[below right, xshift=18, yshift=-10]{$t_{lk}$};
			\node[label= (c)] at (-2,1) {};
		\end{scope}
	\end{tikzpicture}
	\caption{Virtual electron hopping paths between sites lying on a straight line with the same magnitude $|{\bf d}_{ij}|=\dots=|{\bf d}_{jk}|$ . (a) The simplest closed loop hopping path which must be on a line. (b) Closed loop hopping path in the fourth order expansion with same type of bonding for sites lying on a straight line. (c) Closed loop hopping path in the sixth order expansion with same type of bonding for sites lying on a straight line.  Similarly for higher even-order terms. 
    %Same kind of circle is used to indicate that they are of same kind of bonding.
    }
	\label{fig:two-site-closed-loop-hopping-path}
\end{figure}

The results and combinations of virtual hopping paths are more complicated and lengthy at fourth order, whilst the expansion procedure is standard, i.e., same as the procedure in previous orders. We will simply give an example for a particular hopping path, where one can see that only some specific terms are compatible with a dependence on the vector potential of the type $({\bf k}-e{\bf A})$ and $({\bf k}+e{\bf A})$. According to Eq.~\eqref{eq:fourth-order-TDSWT}, the fourth order low-energy effective Hamiltonian reads (we do not need the equation for the fourth order SW generator because we are not going to derive the fifth order result)
\begin{align}
    \tilde{H}^{(4)} 
	&= -\frac{1}{2} \left[i\widehat{S}^{(3)}_{+1}, \widehat{T}_{-1}(t) \right] -\frac{1}{2} \left[i\widehat{S}^{(3)}_{-1}(t), \widehat{T}_{+1}(t)\right] \\     &+ \frac{1}{24} \left[i\widehat{S}^{(1)}_{+1}(t), \left[i\widehat{S}^{(1)}_{-1}(t), \left[i\widehat{S}^{(1)}_{+1}(t), \widehat{T}_{-1}(t)\right]\right]\right] + \frac{1}{24}\left[i\widehat{S}^{(1)}_{+1}(t), \left[i\widehat{S}^{(1)}_{-1}(t), \left[i\widehat{S}^{(1)}_{-1}(t), \widehat{T}_{+1}(t)\right]\right]\right]\nonumber \\
	& +\frac{1}{24}\left[i\widehat{S}^{(1)}_{-1}(t), \left[i\widehat{S}^{(1)}_{+1}(t), \left[i\widehat{S}^{(1)}_{+1}(t), \widehat{T}_{-1}(t)\right]\right]\right] +\frac{1}{24}\left[i\widehat{S}^{(1)}_{-1}(t), \left[i\widehat{S}^{(1)}_{+1}(t), \left[i\widehat{S}^{(1)}_{-1}(t), \widehat{T}_{+1}(t)\right]\right]\right] \nonumber .
\end{align}
After projecting into low-energy subspace and substituting SW generators $\widehat{{S}}^{(3)}(t)$ and $\widehat{{S}}^{(1)}(t)$, the fourth order low-energy effective Hamiltonian reads
\begin{align}
    	\tilde{H}^{(4)}(t) &= -i\iiint dt_3dt_2dt_1 e^{\eta (t_1+t_2+t_3)}\bigg\{-\frac{1}{2}\theta(t-t_1)\theta(t_1-t_2)\theta(t_2-t_3)e^{-iU(t-t_3)} \widehat{T}_{-1}(t)\widehat{T}_0(t_1)\widehat{T}_0(t_2)\widehat{T}_{+1}(t_3)  \nonumber \\
	& +\frac{1}{6}\theta(t-t_1)\theta(t_1-t_2)\theta(t_1-t_3) e^{-iU(t-t_2)}e^{-iU(t_1-t_3)} \widehat{T}_{-1}(t)\widehat{T}_{-1}(t_1)\widehat{T}_{+1}(t_3)\widehat{T}_{+1}(t_2) \nonumber \\
	& -\frac{1}{6}\theta(t-t_1)\theta(t_1-t_2)\theta(t_1-t_3) e^{-iU(t-t_2)}e^{-iU(t_1-t_3)}\widehat{T}_{-1}(t)\widehat{T}_{-1}(t_3)\widehat{T}_{+1}(t_1)\widehat{T}_{+1}(t_2) \nonumber \\
	& + \frac{1}{2}\theta(t-t_1)\theta(t_1-t_2)\theta(t_2-t_3)e^{+iU(t-t_3)} \widehat{T}_{-1}(t_3)\widehat{T}_0(t_2)\widehat{T}_0(t_1)\widehat{T}_{+1}(t)  \nonumber \\
	& -\frac{1}{6}\theta(t-t_1)\theta(t_1-t_2)\theta(t_1-t_3) e^{+iU(t-t_2)}e^{+iU(t_1-t_3)} \widehat{T}_{-1}(t_2)\widehat{T}_{-1}(t_3)\widehat{T}_{+1}(t_1)\widehat{T}_{+1}(t) \nonumber \\
	& +\frac{1}{6}\theta(t-t_1)\theta(t_1-t_2)\theta(t_1-t_3) e^{+iU(t-t_2)}e^{+iU(t_1-t_3)} \widehat{T}_{-1}(t_2)\widehat{T}_{-1}(t_1)\widehat{T}_{+1}(t_3)\widehat{T}_{+1}(t) \nonumber \\
	& - \frac{1}{24}\theta(t-t_1)\theta(t-t_2)\theta(t-t_3) e^{-iU(t-t_1)}e^{+iU(t-t_2)}e^{-iU(t-t_3)}\widehat{T}_{-1}(t_2)\widehat{T}_{-1}(t)\widehat{T}_{+1}(t_3)\widehat{T}_{+1}(t_2) \nonumber \\
	&+ \frac{1}{24}\theta(t-t_1)\theta(t-t_2)\theta(t-t_3) e^{-iU(t-t_1)}e^{+iU(t-t_2)}e^{+iU(t-t_3)}\widehat{T}_{-1}(t_2)\widehat{T}_{-1}(t_3)\widehat{T}_{+1}(t)\widehat{T}_{+1}(t_2) \nonumber \\
	& - \frac{1}{24}\theta(t-t_1)\theta(t-t_2)\theta(t-t_3) e^{+iU(t-t_1)}e^{-iU(t-t_2)}e^{-iU(t-t_3)}\widehat{T}_{-1}(t_1)\widehat{T}_{-1}(t)\widehat{T}_{+1}(t_3)\widehat{T}_{+1}(t_2) \nonumber \\
	& + \frac{1}{24}\theta(t-t_1)\theta(t-t_2)\theta(t-t_3) e^{+iU(t-t_1)}e^{-iU(t-t_2)}e^{+iU(t-t_3)}\widehat{T}_{-1}(t_1)\widehat{T}_{-1}(t_3)\widehat{T}_{+1}(t)\widehat{T}_{+1}(t_2)   \bigg\}.
\end{align}
We can study the general product of the hopping terms
\begin{align}
    \widehat{T}(t)\widehat{T}(t_1)\widehat{T}(t_2)\widehat{T}(t_3) &= \sum_{ijkl} e^{i\left[\theta_{ij}(t)+\theta_{jk}(t_1)+\theta_{kl}(t_2)+\theta_{li}(t_3)\right]} \Bigg\{\sum_{\sigma_1\cdots \sigma_8} c_{i\sigma_1}^\dagger M_{\sigma_1\sigma_2}^{ij} c_{j\sigma_2} c_{j\sigma_3}^\dagger M_{\sigma_3\sigma_4}^{jk}c_{k\sigma_4} \nonumber \\ &\quad \times c_{k\sigma_5}^\dagger M_{\sigma_5\sigma_6}^{kl}c_{l\sigma_6}c_{l\sigma_7}^\dagger M_{\sigma_7\sigma_8}^{li} c_{i\sigma_8}\Bigg\} +\text{h.c.}  \\
    &= \sum_{ijkl} 2\text{Re}\roundbracket{\frac{\tilde{t}_{ij} \tilde{t}_{jk} \tilde{t}_{kl}\tilde{t}_{li}}{U^3}} \bigg\{ ({\bf S}_l\cdot {\bf S}_i+  {\bf S}_k\cdot {\bf S}_i +  {\bf S}_j\cdot {\bf S}_i -  {\bf S}_k\cdot {\bf S}_l -  {\bf S}_j\cdot {\bf S}_l -  {\bf S}_j\cdot {\bf S}_k) \nonumber \\
		&\quad + 4\squarebracket{({\bf S}_j\cdot {\bf S}_k) \left({\bf S}_l\cdot {\bf S}_i\right) - ({\bf S}_j\cdot {\bf S}_l)({\bf S}_k\cdot {\bf S}_i) + ({\bf S}_k\cdot {\bf S}_l)({\bf S}_j\cdot {\bf S}_i)} \bigg\} \nonumber \\
		& + \sum_{ijkl} 2\text{Im} \roundbracket{\frac{\tilde{t}_{ij}\tilde{t}_{jk}\tilde{t}_{kl}\tilde{t}_{li}}{U^3}} \roundbracket{\chi_{kli}+\chi_{jli}+\chi_{jki}-\chi_{jkl}}.
\end{align}
As in the previous case, we have spin terms with real and imaginary parts of a product of hopping amplitudes. Even if we ignore those non-Fleury-Loudon terms, only terms arose from hopping paths with sites lying on a straight line and having the same magnitude of ${\bf d}_{ij}$ could lead to dependence of $({\bf k}-e{\bf A})$ and $({\bf k}+e{\bf A})$ in the low-energy effective spin Hamiltonian. That is, to conclude, the minimal-coupling-like dependence is valid only to second order in $t/U$ for the low-energy effective spin Hamiltonian, and at most for some particular terms in higher order expansions. See Fig.~\ref{fig:two-site-closed-loop-hopping-path} for visualizing the hopping paths with sites lying on a straight line.

%-----------------------------------------------------------------
% Figure: Virtual electron hopping paths in a straight line
%-----------------------------------------------------------------

%-----------------------------------------------------------------
% Two-dimensional honeycomb ferromagnetic topological magnon insulator
%-----------------------------------------------------------------
\section{Two-dimensional honeycomb ferromagnetic topological magnon insulator}\label{sec:2D-FM-magnon-insulator}
In this section, we compute the Berry curvature for the honeycomb ferromagnet (FM) following the standard procedure and derive an analytic relation between RCD and the Berry curvature shown in Eq.~\eqref{eq:RCD-Berry-curvature} in the main text. Recall that after a linearized Holstein-Primakoff (HP) transformation for the spin operators, the magnon Hamiltonian $H^0_M({\bf k})$ can be obtained from the spin Hamiltonian $H^0({\bf k})$ in {\bf k}-space \cite{HP1940}. That is, we have
\begin{equation}
    H^0_M = \frac{1}{2}\sum_{\bf k}\psi_{\bf k }^\dagger H^0_M({\bf k})\psi_{\bf k} \label{eq:magnon_Hamiltonian}
\end{equation}
with the basis $\psi_{\bf k}^\dagger = (a^\dagger_{\bf k},b^\dagger_{\bf k})$, where $a_{\bf k}$ and $b_{\bf k}$ are the HP bosonic operators in $2$ different sublattices. 
For a ferromagnet (FM),  the Berry curvature expression for ferromagnetic magnon bands is the same as the standard form for electronic bands.
Further, a two-band Hermitian Hamiltonian can be represented by a $2\times 2$ Hermitian matrix which admits a decomposition as
\begin{equation}
    H^0_M({\bf k}) = h_0({\bf k}){\bf 1}_2+{\bf h}({\bf k})\cdot \boldsymbol{\tau}, \label{eq:two-band-Hamiltonian}
\end{equation}
where $\boldsymbol{\tau}=(\tau_x,\tau_y,\tau_z)$ is the vector of Pauli matrices, $h_0({\bf k})$ and ${\bf h}({\bf k})=(h_x({\bf k}),h_y({\bf k}),h_z({\bf k}))$ are the corresponding parameters for such decomposition. We can then apply the well-known two-band expression to obtain the Berry curvature~\cite{Niu_2010}
\begin{equation}
    \Omega_\pm({\bf k}) = \mp\frac{1}{2|{\bf h}({\bf k})|^3}{\bf h}({\bf k})\cdot \partial _x{\bf h}({\bf k})\times \partial_y {\bf h}({\bf k}) \label{eq:two-band-expression-Berry-curvature}
\end{equation}
without calculating the eigenenergies and eigenvectors. 
Consider a general Hamiltonian describing a two-dimensional FM in the form of Eq.~\eqref{eq:two-band-Hamiltonian}, the eigenenergies and diagonalization transformation matrix $U_{\bf k}$ are
\begin{align}
    \epsilon_{\pm, \bf k} &= h_0({\bf k})\pm |{\bf h}({\bf k})|,
    \quad U_{\bf k} = (u_{+,\bf k}, u_{-,\bf k}) = \begin{pmatrix}
        \cos \psi_{\bf k} e^{i\varphi_{\bf k}} & -\sin\psi_{\bf k} e^{i\varphi_{\bf k}} \\
        \sin \psi_{\bf k} & \cos \psi_{\bf k} \label{eq:magnon-basis-transformation}
    \end{pmatrix},
\end{align}
where $u_{\pm,\bf k}$ are the eigenvectors that correspond to $\epsilon_{\pm,\bf k}$. The trigonometric parameters are defined as
\begin{equation}
    \cos \psi_{\bf k} = \sqrt{\frac{|{\bf h}({\bf k})|+h_z({\bf k})}{2|{\bf h}({\bf k})|}}, \quad \sin \psi_{\bf k} =\sqrt{\frac{|{\bf h}({\bf k})|-h_z({\bf k})}{2|{\bf h}({\bf k})|}}, \quad
    e^{-i\varphi_{\bf k}} = \frac{1}{\sqrt{h_x^2({\bf k})+h_y^2({\bf k})}}(h_x({\bf k})+ih_y({\bf k})) = \frac{r_{\bf k}}{|r_{\bf k}|}, \label{eq:parameters-two-band-diagonalization}
\end{equation}
where $r_{\bf k}=h_x({\bf k})+ih_y({\bf k})$. With these equations, we can derive the general relation between RCD and Berry curvature in FMs once the Hamiltonian in Eq.~\eqref{eq:two-band-expression-Berry-curvature} satisfies the condition in Eq.~\eqref{eq:even-combination-of-k-eA-suppl}.

\subsection{RCD and Berry curvature in FM}
Applying the shortcut as per Eq.~\eqref{eq:FL_momentum_derivative} in the main text, for a two-band ferromagnetic magnon Hamiltonian in ${\bf k}$-space (Eq.~\eqref{eq:two-band-Hamiltonian}) that fulfills the condition Eq.~\eqref{eq:even-combination-of-k-eA-suppl}, which we have proved its validity in the previous section, the FL vertex for circularly polarized light is obtained as
\begin{equation}
    H^{ss'}_{R,{\bf k}} = \frac{K_0}{2} \hat{O}^{ss'}(h_0+{\bf h}\cdot\boldsymbol{\sigma}), \label{eq:FL-vertex}
\end{equation}
where we defined a differential operator
\begin{equation}
    	\hat{O}^{ss'} = \partial_x^2 + ss'\partial_y^2 + i(s-s')\partial_x \partial_y = \hat{A}^{ss'}+i\hat{B}^{ss'}
\end{equation}
such that $\hat{O}^{RR}=\hat{O}^{LL} = \partial_x^2+\partial_y^2=\hat{A}^{RR}$, $\hat{O}^{RL} = \partial_x^2-\partial_y^2-2i\partial_x\partial_y = \hat{A}^{RL}+i\hat{B}^{RL}$ and $\hat{O}^{LR}=(\hat{O}^{RL})^*=\hat{A}^{RL}-i\hat{B}^{RL}$ with $\hat{A}^{ss'}=\hat{A}^{s's}$ and $\hat{B}^{RL}=-\hat{B}^{LR}$. 
Here, we have dropped the explicit dependence on {\bf k} for brevity, $s=\pm 1$ denotes left (L) and right (R) circularly polarized light and $K_0=\frac{2e^2}{\hbar^2(U-\omega_{\text{in}})}A_{\text{sc}} A_{\text{in}}$.
We then denote ${\bf A}^{ss'}=(A^{ss'}_x,A_y^{ss'},A_z^{ss'})=\hat{A}^{ss'}{\bf h}$ and ${\bf B}^{ss'}=(B^{ss'}_x,B_y^{ss'},B_z^{ss'})=\hat{B}^{ss'}{\bf h}$. The diagonal term is invariant under the basis transformation, and hence we denote $A_0^{ss'}=\hat{A}^{ss'}h_0$ and $B^{ss'}_0=\hat{B}^{ss'}h_0$. Then, the FL vertex in Eq.~\eqref{eq:FL-vertex} in terms of $A$ and $B$ reads
\begin{equation}
	H_{R,\bf k}^{ss'} = \frac{K_0}{2}\left[A_0^{ss'}+iB^{ss'}_0+({\bf A}^{ss'}+i {\bf B}^{ss'})\cdot \boldsymbol{\sigma}\right].
\end{equation}
We first note several identities before moving on. In the magnon basis, the diagonalization means that $u_\pm^\dagger ({\bf h}\cdot \boldsymbol{\sigma})u_\pm = h_0\pm |\bf h|=\epsilon_{\pm,\bf k}$. Hence, we have the identities
\begin{equation}
    u_+^\dagger (\boldsymbol{\sigma})u_+ = \frac{{\bf h}}{|{\bf h}|} = {\bf n} \quad \text{and} \quad u_-^\dagger (\boldsymbol{\sigma})u_- = -\frac{{\bf h}}{|{\bf h}|} = -{\bf n}.
\end{equation}
Furthermore, we define a complex-valued vector as
\begin{equation}
    \boldsymbol{\Lambda} =  u_+^\dagger (\boldsymbol{\sigma})u_- \quad \text{and} \quad  \boldsymbol{\Lambda}^* =u_-^\dagger (\boldsymbol{\sigma})u_+,
\end{equation}
which is orthogonal to ${\bf h}$ in the sense of ${\bf h}\cdot \boldsymbol{\Lambda}=0$. Then the cross product of its complex-conjugated vector will be the normal vector $(\boldsymbol{\Lambda}\times \boldsymbol{\Lambda}^*) = 2i{\bf n}$. We are interested in only interband transitions for an FM, which are given by the transition amplitude
\begin{equation}
    u_+^\dagger H_{R,{\bf k}}^{ss'} u_- = \frac{K_0}{2}({\bf A}^{ss'}+i {\bf B}^{ss'})\cdot \boldsymbol{\Lambda}.
\end{equation}
Since we are computing the difference of the square of the transition amplitude for RCD,
the terms corresponding to $s=s'$ will be equal and cancel, that is,
\begin{equation}
    \chi_{\bf k} = \sum_{s'} \left\vert u_+^\dagger H_{R,\bf k}^{Rs'} u_{-}\right\vert^2 - \left\vert u_+^\dagger H_{R,\bf k}^{Ls'} u_{-}\right\vert^2 =\left\vert u_+^\dagger H_{R,\bf k}^{RL'} u_{-}\right\vert^2- \left\vert u_+^\dagger H_{R,\bf k}^{LR} u_{-}\right\vert^2. \label{eq:two-band-RCD-interband}
\end{equation}
Hence, it suffices to compute the case $s\neq s'$, which gives
\begin{equation}
    u_+^\dagger H_{R,{\bf k}}^{RL} u_- = \frac{K_0}{2}({\bf A}^{RL}+i {\bf B}^{RL})\cdot \boldsymbol{\Lambda} \quad \text{and} \quad u_+^\dagger H_{R,{\bf k}}^{LR} u_- = \frac{K_0}{2}({\bf A}^{RL}-i {\bf B}^{RL})\cdot \boldsymbol{\Lambda}. 
\end{equation}
Therefore, the terms in Eq.~\eqref{eq:two-band-RCD-interband} can be re-written as
\begin{align}
    \left\vert u_+^\dagger H_{R,\bf k}^{RL} u_{-}\right\vert^2 &= \frac{K_0^2}{4}\squarebracket{|{\bf A}^{RL}\cdot \boldsymbol{\Lambda}|^2+ |{\bf B}^{RL}\cdot \boldsymbol{\Lambda}|^2 + i({\bf A}^{RL}\cdot \boldsymbol{\Lambda}^*)({\bf B}^{RL}\cdot \boldsymbol{\Lambda})-i({\bf A}^{RL}\cdot \boldsymbol{\Lambda})({\bf B}^{RL}\cdot \boldsymbol{\Lambda}^*)} \\
   \left\vert u_+^\dagger H_{R,\bf k}^{LR} u_{-}\right\vert^2 &= \frac{K_0^2}{4}\squarebracket{|{\bf A}^{RL}\cdot \boldsymbol{\Lambda}|^2+ |{\bf B}^{RL}\cdot \boldsymbol{\Lambda}|^2 +i({\bf A}^{RL}\cdot \boldsymbol{\Lambda})({\bf B}^{RL}\cdot \boldsymbol{\Lambda}^*) - i({\bf A}^{RL}\cdot \boldsymbol{\Lambda}^*)({\bf B}^{RL}\cdot \boldsymbol{\Lambda})}.
\end{align}
The RCD becomes $\chi_{\bf k}=\frac{iK_0^2}{2} \left[ ({\bf A}^{RL}\cdot \boldsymbol{\Lambda}^*)({\bf B}^{RL}\cdot \boldsymbol{\Lambda})-({\bf A}^{RL}\cdot \boldsymbol{\Lambda})({\bf B}^{RL}\cdot \boldsymbol{\Lambda}^*) \right]$. Using the standard vector algebra identity $({\bf a \times b})\cdot ({\bf c\times d})= ({\bf a \cdot c})({\bf b \cdot d}) - ({\bf a\cdot d})({\bf b \cdot c})$ for some vectors ${\bf a},{\bf b},{\bf c}$ and ${\bf d}$, the RCD can be written as a scalar triple product similar to the two-band expression of the Berry curvature in Eq.~\eqref{eq:two-band-expression-Berry-curvature}
\begin{equation}
    \chi_{\bf k} = K_0^2 {\bf n}\cdot ({\bf A}^{RL}\times {\bf B}^{RL}). \label{eq:two-band-RCD-triple-scalar-product}
\end{equation}
In order to extract the relation between the RCD and the Berry curvature, we re-write the vector ${\bf h}$ as $|{\bf h}|{\bf n}$ in vectors ${\bf A}^{RL}$ and ${\bf B}^{RL}$
\begin{align}
    {\bf A}^{RL} &= (\partial_x^2-\partial_y^2)(|{\bf h}|{\bf n}) 
    =(\partial_x^2|{\bf h}|-\partial_y^2|{\bf h}|) {\bf n} + |{\bf h}|(\partial_x^2{\bf n} - \partial_y^2{\bf n}) + 2(\partial_x |{\bf h}|)(\partial_x{\bf n }) - 2(\partial_y |{\bf h}|)( \partial_y {\bf n}) \label{eq:two-band-vector-A-RL} \\
    {\bf B}^{RL} &= (-2\partial_x\partial_y)(|{\bf h}|{\bf n}) = -2(\partial_x\partial_y |{\bf h}|){\bf n} -2|{\bf h}|(\partial_x\partial_y {\bf n}) - 2(\partial_x |{\bf h}|)(\partial_y {\bf n}) - 2(\partial_y |{\bf h}|)(\partial_x {\bf n}). \label{eq:two-band-vector-B-RL}
\end{align}
We can use ${\bf n}\cdot {\bf n}=1$ to re-write the second derivative of ${\bf n}$ in terms of first derivative by the quantum metric
\begin{equation}
    {\bf n}\cdot \partial_\mu {\bf n } = 0 \Rightarrow  4g_{\mu\nu} = \partial_\mu {\bf n} \cdot \partial_\nu {\bf n}= -{\bf n}\cdot \partial_\mu \partial_\nu {\bf n }. \label{eq:identity-quantum-metric-normalized-vector}
\end{equation}
Next, with the help of differential geometry, we use that the second derivative of ${\bf n}$ can be decomposed into a tangential part with the Christoffel symbol $\Gamma^\lambda_{\mu\nu}$ and a normal part with the quantum metric
\begin{equation}
    \partial_\mu\partial_\nu {\bf n} = \sum_\lambda \Gamma^\lambda_{\mu\nu} \partial_\lambda {\bf n}  - 4g_{\mu\nu} {\bf n}. \label{eq:second-derivative-of-normalized-vector}
\end{equation}
Using Eqs.~\eqref{eq:identity-quantum-metric-normalized-vector} and \eqref{eq:second-derivative-of-normalized-vector}, we can see that the contribution of quantum metric is stored in the Christoffel symbol as
\begin{equation}
    \Gamma_{\lambda \mu\nu} = \sum_{\rho}g_{\lambda \rho} \Gamma^\rho_{\mu\nu} = 2\left( \partial_\mu g_{\nu\lambda} +\partial_\nu g_{\lambda \mu} -  \partial_\lambda g_{\mu\nu} \right). \label{eq:Christoffel-symbol-and-quantum-metric}
\end{equation}
Eventually, using Eqs.~\eqref{eq:two-band-RCD-triple-scalar-product}, \eqref{eq:two-band-vector-A-RL}, \eqref{eq:two-band-vector-B-RL} and \eqref{eq:second-derivative-of-normalized-vector}, we arrive at a clean relation between the RCD and Berry curvature
\begin{align}
    \chi_{\bf k}  &= \pm K_0^2 \Omega_\pm ({\bf k}) \{ 2\left\vert \nabla (\Delta_{\bf k}) \right\vert^2 + \Delta_{\bf k}^2  \squarebracket{(\Gamma_{xx}^x-\Gamma_{yy}^x)\Gamma_{xy}^y  -(\Gamma_{xx}^y-\Gamma_{yy}^y)\Gamma_{xy}^x  } \nonumber \\
    &\quad +\Delta_{\bf k}\left [(\partial_x\Delta_{\bf k}) (\Gamma^x_{xx}-\Gamma^x_{yy}+2\Gamma_{xy}^x)-(\partial_y\Delta_{\bf k})(\Gamma^y_{xx}-\Gamma^y_{yy}-2\Gamma_{xy}^y)\right], \label{eq:two-band-RCD-Berry-cruvature} \\
    &= \pm \Omega_{\pm}({\bf k}) \varrho_{\bf k}
\end{align}
where $\Delta_{\bf k} = \epsilon_{+,\bf k}-\epsilon_{-,\bf k}$ is the energy gap and $\nabla = \partial_x^2+\partial_y^2$ is the gradient operator in ${\bf k}$-space.
The RCD is indeed directly proportional to the Berry curvature in momentum space with a weighting function $\varrho_{\bf k}$ related to the local geometry of the energy gap, i.e.,
\begin{align}
    \varrho_{\bf k}/K_0^2 &= 2\left\vert \nabla (\Delta_{\bf k}) \right\vert^2 + \Delta_{\bf k}^2  \squarebracket{(\Gamma_{xx}^x-\Gamma_{yy}^x)\Gamma_{xy}^y  - (\Gamma_{xx}^y-\Gamma_{yy}^y)\Gamma_{xy}^x  } \nonumber \\
    &\quad +\Delta_{\bf k}\left [(\partial_x\Delta_{\bf k}) (\Gamma^x_{xx}-\Gamma^x_{yy}+2\Gamma_{xy}^x)-(\partial_y\Delta_{\bf k})(\Gamma^y_{xx}-\Gamma^y_{yy}-2\Gamma_{xy}^y)\right].
\end{align}
This connection is more general in the sense that whenever we can replace the vector potential derivatives by the momentum derivatives of a two-band Hermitian Hamiltonian in momentum space at second order, all steps should follow analogously.

%-----------------------------------------------------------------
% An example of 2D Ferromagnet on the honeycomb lattice
%-----------------------------------------------------------------
\section{An example of 2D Ferromagnet on the honeycomb lattice}\label{sec:example-2D-FM-honeycomb-CrI3}
Here, we show explicitly the workflow of the established machinery in the main text and in the previous section with the spin model of $\textrm{CrI}_3$ obtained from Ref.~\cite{Chen2018}, which reads
\begin{equation}
    \label{eq:H0_FM_SS_suppl}
    H^0 = -\sum_{r=1}^3\sum_{\langle i,j\rangle_r} J_r {\bf S}_i\cdot {\bf S}_j - \sum_{\langle i,j\rangle_2} {\bf D}_{ij} \cdot ({\bf S}_i\times {\bf S}_j) -\sum_{i} A_z (S_i^z)^2 ,
\end{equation}
where $\langle i,j\rangle_r$ denotes the sites between $r^\mathrm{th}$ nearest neighbors (n.n.). $A_z$ is the easy-axis anisotropy, $J_r$ is the exchange coefficient between the $r^\mathrm{th}$ n.n. and ${\bf D}_{ij}$ is the DMI vector between sites $i$ and $j$. The corresponding magnon model in the representation of Eq.~\eqref{eq:two-band-Hamiltonian} reads
\begin{equation}
    h_0 = M+2A_zS-\gamma_{2,\bf k}, \quad h_x = -\text{Re} \left( \gamma_{1, \bf k} + \gamma_{3, \bf k} \right), \quad h_y= \text{Im} \left(\gamma_{1, \bf k} + \gamma_{3, \bf k}  \right), \quad h_z= \lambda_\textbf{k},
\end{equation}
where $M=3J_1S+6J_2S+3J_3S$, $\gamma_{1, \textbf{k}} = J_1S\sum_{\langle i,j\rangle_1} e^{i\textbf{k}\cdot \boldsymbol{\delta}^{(1)}}$, $\gamma_{2,\textbf{k}} = 2J_2S\sum_{\langle ij\rangle_2} \cos(\textbf{k}\cdot \boldsymbol{\delta}^{(2)})$, $\gamma_{1, \textbf{k}} = J_3S\sum_{\langle i,j\rangle_3} e^{i\textbf{k}\cdot \boldsymbol{\delta}^{(3)}}$ and $\lambda_\textbf{k} = 2DS\sum_{\langle ij\rangle_2} \sin (\textbf{k} \cdot \boldsymbol{\delta}^{(2)})$. With the vector ${\bf h}$, one can obtain the magnon bands, eigenvectors, Berry curvature, RCD in momentum space immediately using Eqs.~\eqref{eq:two-band-expression-Berry-curvature},~\eqref{eq:magnon-basis-transformation},~\eqref{eq:parameters-two-band-diagonalization} and \eqref{eq:two-band-RCD-Berry-cruvature}. 

The second-order LMCs are given by 
\begin{equation}
L_{\mu\nu}^{(2)}({\bf k})=\frac{1}{2}\partial_\mu\partial_\nu \begin{pmatrix}
        \gamma_{2,\bf k} + d_{z,\bf k} &  \gamma_{1,\bf k}+\gamma_{3,\bf k} \\ \gamma^*_{1,\bf k} + \gamma^*_{3,\bf k} &  \gamma_{2,\bf k} -d_{z,\bf k}
    \end{pmatrix}\,,
\end{equation}
where we remind the reader that the derivatives are taken with respect to the components of the momentum ${\bf k }$. 

\end{document}